UNIVERSIDADE FEDERAL DO RIO DE JANEIRO
CENTRO DE CIÊNCIAS MATEMÁTICAS E DA NATUREZA
OBSERVATÓRIO DO VALONGO
DEPARTAMENTO DE ASTRONOMIA

# "Análise de uma nova série de medidas de variação do semidiâmetro do Sol"

Aluno: Sérgio Calderari Boscardin

Orientador: Alexandre Humberto Andrei

(Observatório Nacional / Observatório do Valongo)

Projeto Final de Curso para a Obtenção do Título de Astrônomo

Rio de Janeiro - Maio 2004

**O Sol visto em 12 de março de 2004 com uma mancha maior que a Terra.**

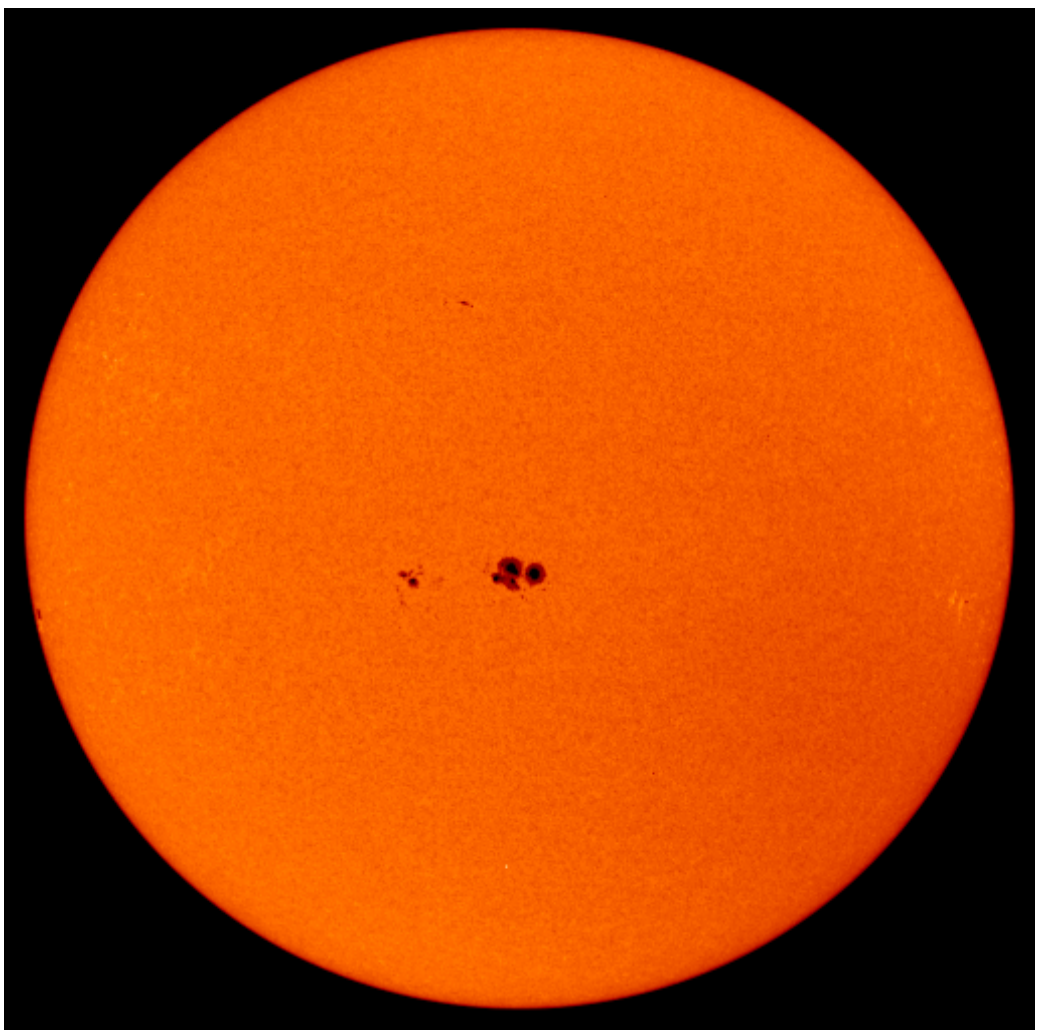

"Foi ele que me deu a verdadeira ciência das coisas que existem, para que eu conheça a constituição do universo, e as propriedades dos elementos. O princípio, o fim e o meio dos tempos, as mudanças dos solstícios e as vicissitudes das estações, os cursos do ano, e as disposições das estrelas,.... aprendi tudo que há escondido e não descoberto, porque a sabedoria que tudo criou mo ensinou."

Trecho do livro da Sabedoria escrito no Egito, talvez em Alexandria entre os anos 150 e 50 a.C.

**RESUMO.**

Desde 1997 o Observatório Nacional vem observando o Sol com um astrolábio com o objetivo de estudar as variações do semidiâmetro solar. Até o momento mais de 20.000 observações foram feitas. As observações efetuadas durante o ano de 2001 são interessantes pois, pela primeira vez, podem ser comparadas com outras séries equivalentes disponíveis neste ano e pelo fato do Sol passar por um máximo recorrente de seu ciclo de atividades neste mesmo ano, o que pode ser usado para se correlacionar o semidiâmetro e o ciclo de atividades do Sol.

Os dados observados foram inicialmente selecionados, retirando-lhes uma série de valores referentes a observações mal sucedidas. A seguir foram retirados erros introduzidos nos valores observados decorrentes da instabilidade do prisma objetivo. Esta instabilidade tende a aumentar os valores observados a leste e a diminuir os observados a oeste. Assim, foi possível retirar este erro ao se comparar as duas séries, ou seja, a leste e a oeste. A seguir foram selecionados, entre outros, três parâmetros de observação que introduzem erros nos valores de semidiâmetro solar. Uma vez calculada sua influência, esta foi retirada da série. Finalmente foi retirado da série o erro decorrente de um desvio da vertical do eixo de rotação azimutal do astrolábio e introduzido pelo azimute de observação. Todas as correções efetuadas são bem menores que os desvios padrão da série e as médias permaneceram inalteradas o que garante a integridade da série. Após a correção da instabilidade do prisma objetivo as séries a leste e a oeste passaram a ter o mesmo comportamento sinalizando desta forma que são séries idênticas relativas ao mesmo objeto de estudo.

A série corrigida foi, então, comparada com as observações do semidiâmetro solar obtidas pelo CERGA, do Observatório da Cote d'Azur na França e pelas obtidas pelo Observatório de Tubitak em Antalya na Turquia. A comparação revela semelhanças que merecem ser estudadas em maior detalhe. A série foi também comparada com o número médio diário de manchas solares que tem três máximos durante o ano de 2001, coincidentes com três máximos da série corrigida desde que a série seja atrasada de 50 dias. Isto é indicativo para que se proceda alguma investigação mais aprofundada.

**ABSTRACT.**

From 1997 the Observatório Nacional has been observing the Sun with an astrolabe in order to study variations on the solar semi-diameter. Till now more than 20.000 observations were done. The observations taken in the year of 2001 were special because, for the first time, they can be compared with other equivalent series, and because the Sun passed by a second maximum of its cycle of activity.

The observed data were at first selected, by leaving out the values relative to unsuccessful measurements. We dealt with the errors in the observed values originating from the instability of the objective prism, which increases the east values and diminishes the west ones. Then, it was possible to remove the errors by comparing the east and west series. Next, the observational conditions were investigated: the air temperature, the Fried's parameter, and the dispersion on the solar limb adjustment. Their influence on the results was accounted for. Finely we considered the errors arising from leveling defects, modeled by a dependence to the azimuth of observation. All the corrections are smaller than the standard deviation and the mean semi-diameter values stay unaffected. This means that the series have the same pattern with smaller deviations. After the correction of the prism instability the east and west series become more similar, what reinforces that they aim at the same target.

The corrected series were compared with other observations of solar semi-diameter from the CERGA, Côte d'Azur Observatory in France and from Tubitak Observatoty in Antalya, Turkey. The comparison reveals alike features, which deserve further, detailed study. Also, the series was compared to the daily mean Sun spots number. It exhibited three maxima along 2001, that coincide on time with three maxima of the correct series delayed by 50 days. Again, this requires a further study.

# LISTA DE FIGURAS.

**LISTA DE TABELAS.**

**INTRODUÇÃO.**

O diâmetro do Sol tem sido medido ao longo de toda a história. Arquimedes (287-212 AC) teria calculado um valor entre 27 e 33 minutos de arco. Aristarco (310-230 AC) atribuiu o valor de 30 minutos e Ptolomeu (87-151 DC) o valor de 31'20". Ptolomeu acompanhou a medida por um ano buscando verificar as variações sem contudo percebê-las. Em 1951, Tycho Brahe (1546-1601) realizou onze medidas que segundo Johannes Kepler (1571-1630) mostram um valor mínimo do diâmetro solar de 30'30". Dados históricos apontam para um raio solar maior durante o período conhecido como Mínimo de Maunder. Em 1891 Auwers obteve um valor de 959",53 para o semidiâmetro do Sol. [1]

Muitos têm se preocupado com a medida do diâmetro solar e com as suas possíveis variações e, na lista das instituições que efetivaram algum programa neste sentido está o Observatório Nacional – ON que modificou em 1997 um Astrolábio Danjon, na sua sede no Rio de Janeiro, equipando-o com um prisma refletor de ângulo variável e uma câmara CCD, o que permitiu o monitoramento do diâmetro solar[2]. Desde aquela data um vasto programa de observação e medida do diâmetro solar vem sendo desenvolvido e um total de mais de 20.000 observações está arquivado.

No ano de 2001 o Sol passou por um máximo de atividade de seu atual ciclo de manchas de onze anos. Se considerarmos a média mensal de manchas solares o máximo de atividades do Sol foi caracterizado por dois picos bem distintos, o primeiro deles culminando em julho de 2000 e o segundo em setembro de 2001 havendo entre os dois picos um declínio de atividades. A Figura I mostra a média mensal de manchas dos últimos cinco ciclos. O número médio mensal de manchas solares durante os meses do ano de 2001 é mostrado na Tabela I, a seguir.

**Tabela I – Número médio mensal de manchas solares durante o ano de 2001.**

| JAN | FEV | MAR | ABR | MAI | JUN | JUL | AGO | SET | OUT | NOV | DEZ |
|-----|-----|-----|-----|-----|-----|-----|-----|-----|-----|-----|-----|
| 95,6 | 80,6 | 113,5 | 107,7 | 96,6 | 134,0 | 81,8 | 106,4 | 150,7 | 125,7 | 106,5 | 132,2 |

Estes números devem ser comparados com a média dos valores a longo termo. Os valores da Figura I cujo período vai de janeiro de 1955 a dezembro de 2003 têm por média 78,6.

**Figura I - Médias mensais dos números de manchas
do Sol de janeiro de 1955 a janeiro de 2004.**

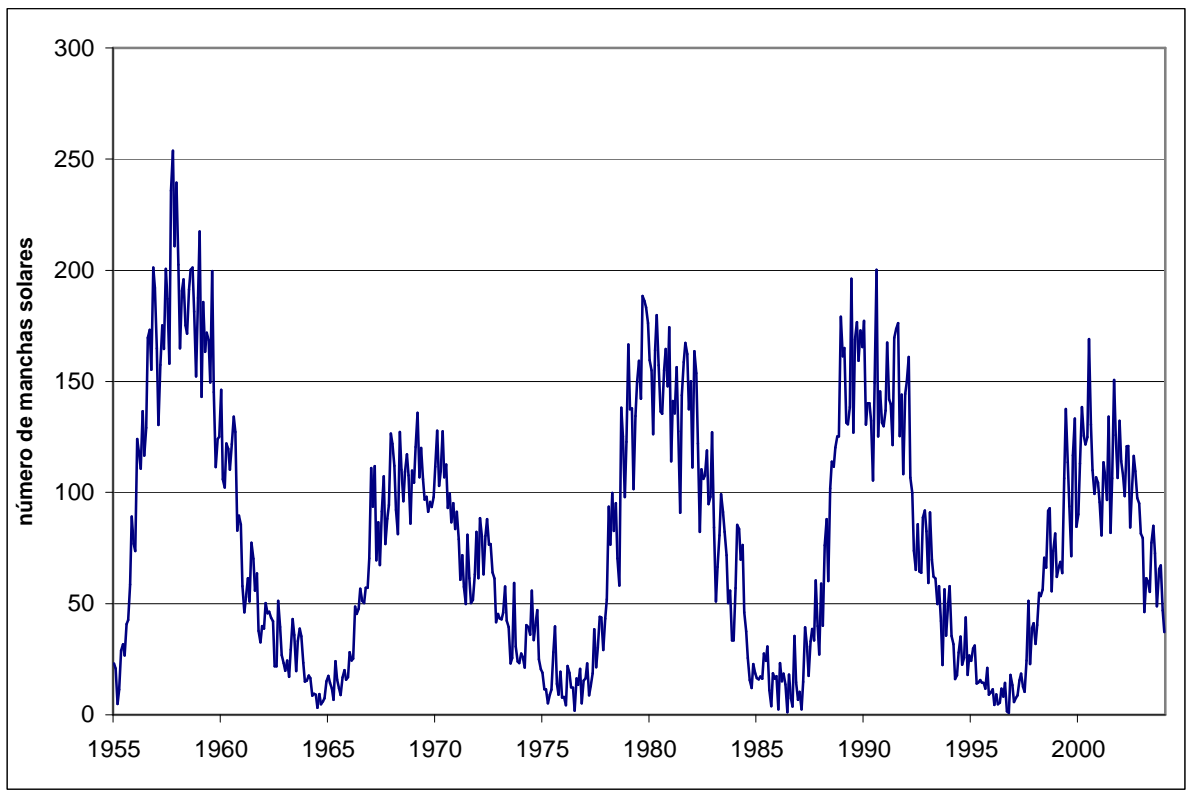

Assim, avaliar as variações do diâmetro solar no ano de 2001 nos dá uma excelente oportunidade de compará-las com a evolução das atividades do Sol que são particularmente diferentes neste ano de 2001.

Uma segunda razão para a escolha de 2001 é que neste ano ocorre pela primeira vez a disponibilidade de dados obtidos por outras equipes. São dados obtidos com instrumentos e métodos análogos, porém observando outras heliolatitudes e com condições atmosféricas diferentes. Podemos então comparar nossos dados com outros obtidos no CERGA na França e também em Antalya na Turquia.

Para podermos efetivar a comparação dos dados do ON com os dados de atividade solar e também com os dados de outros observatórios é necessário retirar dos dados observados toda uma série de erros decorrentes da observação ou introduzidos pelo instrumento. Sabe-se que o não exato nivelamento do instrumento induz um pequeno erro aos valores observados, assim como a temperatura do ar no instante da observação. O conjunto de molas que mantém a orientação do prisma objetivo introduz também um pequeno erro já que este não permanece estável durante o espaço de tempo em que ocorre a observação.

Detectar estas influências, a magnitude dos erros introduzidos, e, finalmente, corrigir os valores observados, retirando deles estes erros, é a tarefa que nos dispusemos a fazer e que está no escopo deste trabalho.

# O SEMIDIÂMETRO SOLAR.

Inicialmente um esclarecimento de porque se fala em semidiâmetro do Sol e não em raio. Historicamente os observadores se referiam ao raio do Sol, pois, imaginavam-no perfeitamente circular e, calculavam-lhe o raio. Mais recentemente, quando se percebeu que sua figura não era circular e portanto, o diâmetro dependia da latitude medida, passaram a medir-lhe os diâmetros e por comparação com valores anteriormente determinados, dividiam-no por dois para comparar com o raio e assim passou-se a atribuir valores ao que hoje designamos por semidiâmetro do Sol.

O Sol é um corpo fluido em rotação. Esta rotação deve provocar algum efeito na forma. Há também, pequenas depressões na superfície solar em torno das regiões conhecidas como 'royal zones' [3]. Estes fatos apontam para uma forma não esférica para o Sol.

O Sol é uma estrela da seqüência principal e assim, de uma maneira simplificada, sua estrutura se mantém equilibrada por duas forças. De um lado a pressão interna tende a aumentar-lhe o volume empurrando as camadas exteriores para fora. Contrapondo-se a esta, a atração gravitacional tende a diminuir o volume solar, puxando todas as camadas para o centro. As forças de pressão são alimentadas pela intensa temperatura interna que é constantemente mantida pelas reações nucleares que ocorrem no interior, uma vez que a energia interna flui para o exterior sendo liberada na superfície exterior sob a forma de radiação.

Durante toda a vida do Sol na seqüência principal este quadro praticamente não se altera, mas podem ocorrer pequenos desvios do equilíbrio, e, qualquer pequeno desvio no fluxo energético para o exterior ou mesmo na produção de energia interna é rapidamente acompanhado por uma nova acomodação das camadas que pode resultar em pequenas alterações do volume solar. Fenômenos de ordem magnética que ocorrem nas camadas mais externas podem também alterar o volume destas camadas sujeitas a pressões menores. Assim, podemos esperar que o semidiâmetro solar possa sofrer pequenas alterações, que podem ser totalmente aleatórias ou atender a algum tipo de lei cíclica.

Estudos recentes de heliossismologia mostram que o formato do Sol varia durante o ciclo solar [4].

O diâmetro solar pode variar no tempo e também ao longo de sua forma. Como varia o diâmetro solar ao longo de sua forma? Como varia o diâmetro solar ao longo do tempo? Para responder a estas questões, o Observatório Nacional, implementou em 1997 seu programa de observação do diâmetro solar.

**O ASTROLÁBIO SOLAR.**

No Observatório Nacional o Sol vem sendo observado e seu semidiâmetro medido desde 1997. Mais de 20.000 observações foram feitas desde aquele ano até atualmente. A Tabela II a seguir mostra o número de observações a cada ano, bem como o número de dias em que se observou a cada ano.

**Tabela II – Número de observações do semidiâmetro do Sol
e o número de dias em que se observou de 1997 a 2003. [ 12]**

| ANO | 1997 | 1998 | 1999 | 2000 | 2001 | 2002 | 2003 | Total |
|---|---|---|---|---|---|---|---|---|
| Número de observações | 2.706 | 3.927 | 3.949 | 3.268 | 1.890 | 2.905 | 2.127 | 20.774 |
| Número de Dias | 158 | 162 | 157 | 163 | 122 | 154 | 134 | 1.050 |

Numa rotina funcional, quando há condições de observação, são feitas seções de medição do semidiâmetro solar. Parte das seções são feitas antes da passagem meridiana do Sol e parte após sua passagem meridiana. Durante o ano de 2001 o semidiâmetro solar foi assim medido mais de 1892 vezes. Pode-se ver no quadro que o ano de 2001 foi o ano de menor número de observações, assim como o de menor número de dias em que se observou. Isto se deve ao fato do astrolábio ter sido desativado no final de setembro para sofrer uma manutenção, retornado ao serviço apenas nos últimos dias do ano.

O astrolábio solar do Observatório Nacional - ON consiste de um telescópio refrator na frente do qual é colocado um prisma refletor e uma bacia com mercúrio. O prisma tem duas faces refletoras que fazem o mesmo ângulo com um plano horizontal. Os raios incidentes são separados em dois feixes: um devido à reflexão na face superior do prisma e o outro obtido por reflexão na face inferior do prisma após uma primeira reflexão na face de mercúrio. Nestas condições obtêm-se duas imagens do objeto observado que tem o mesmo deslocamento horizontal, mas deslocamentos verticais opostos. Assim, o instrumento pode detectar a coincidência das duas imagens do mesmo objeto, quando este cruza uma linha de distância zenital determinada pelo ângulo do prima. A Figura II mostra um esquema do Astrolábio solar.

**Figura II - Esquema do prisma e da bacia com mercúrio do Astrolábio, indicando o caminho que fazem os raios do Sol até chegarem ao detector.**

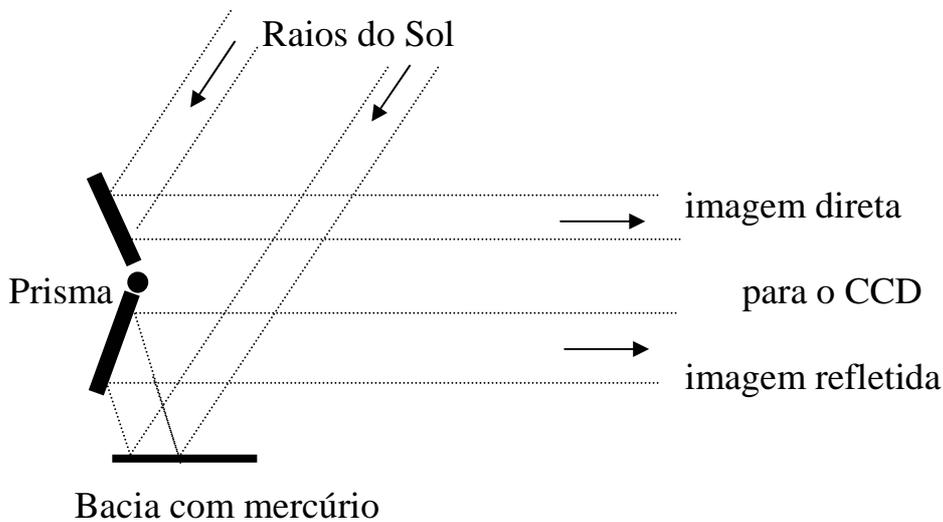

Fazendo-se variar o ângulo com o plano horizontal, simultaneamente, das duas faces do prisma, obtêm-se imagens de objetos em diferentes alturas zenitais. O instrumento instalado no ON permite a observação de objetos entre 25º e 55º de distância zenital.

O astrolábio fornece a uma câmara CCD as duas imagens do Sol. A cada observação de semidiâmetro solar escolhe-se determinada distância zenital, por onde deve passar o Sol. Um círculo de distância zenital constante é conhecido como almicantarado. A medida em que o Sol se aproxima do almicantarado escolhido, as duas imagens se aproximam e, quando o primeiro bordo do Sol cruza o almicantarado, as duas imagens se tocam. Quando o segundo bordo do Sol cruza esta linha, as imagens se separam.

Quando o bordo do Sol está para cruzar o almicantarado, são feitas 46 imagens. O instante em que cada imagem é obtida é fornecido pelo relógio atômico do ON. A análise destas imagens fornece o exato instante em que as duas imagens, a direta e a refletida, se tocaram ou se separaram.

Conhecendo-se a marcha do Sol a cada dia do ano e a cada posição que ocupa na esfera celeste, pode-se calcular, a partir do tempo que o Sol levou para cruzar totalmente o almicantarado, o seu tamanho angular. Este tamanho é então reduzido para a distância média do Sol, isto é, para uma UA, obtendo-se a medida angular do seu diâmetro.

As imagens obtidas são dirigidas ao CCD que tem 512 linhas e 512 colunas de pixels. Por conta do entrelaçamento das imagens, tomam-se apenas 256 linhas. Um pixel corresponde a 0",56 e desta forma apenas uma parte do bordo solar é tomada na imagem. Em cada imagem são identificados 256 pontos do bordo direto e 256 pontos do bordo refletido do Sol, um para cada uma das linhas da imagem. As linhas apresentam uma curva de intensidade de luz. Esta curva tem a propriedade de apresentar um extremo de sua derivada ao longo do bordo solar. A segunda derivada da função de intensidade de luz é igual a zero no bordo solar. Assim, o bordo solar é o ponto em que a curva de luz tem seu ponto de inflexão. Na Figura III há uma esquema da determinação de um ponto do bordo solar. Na verdade a análise é um pouco mais complexa porque há dois bordos na mesma imagem, o direto e o refletido [5].

**Figura III - Esquema da determinação de um ponto do bordo solar. A curva de luz ao longo de uma coluna do CCD é ajustada e determina-se seu ponto de inflexão**

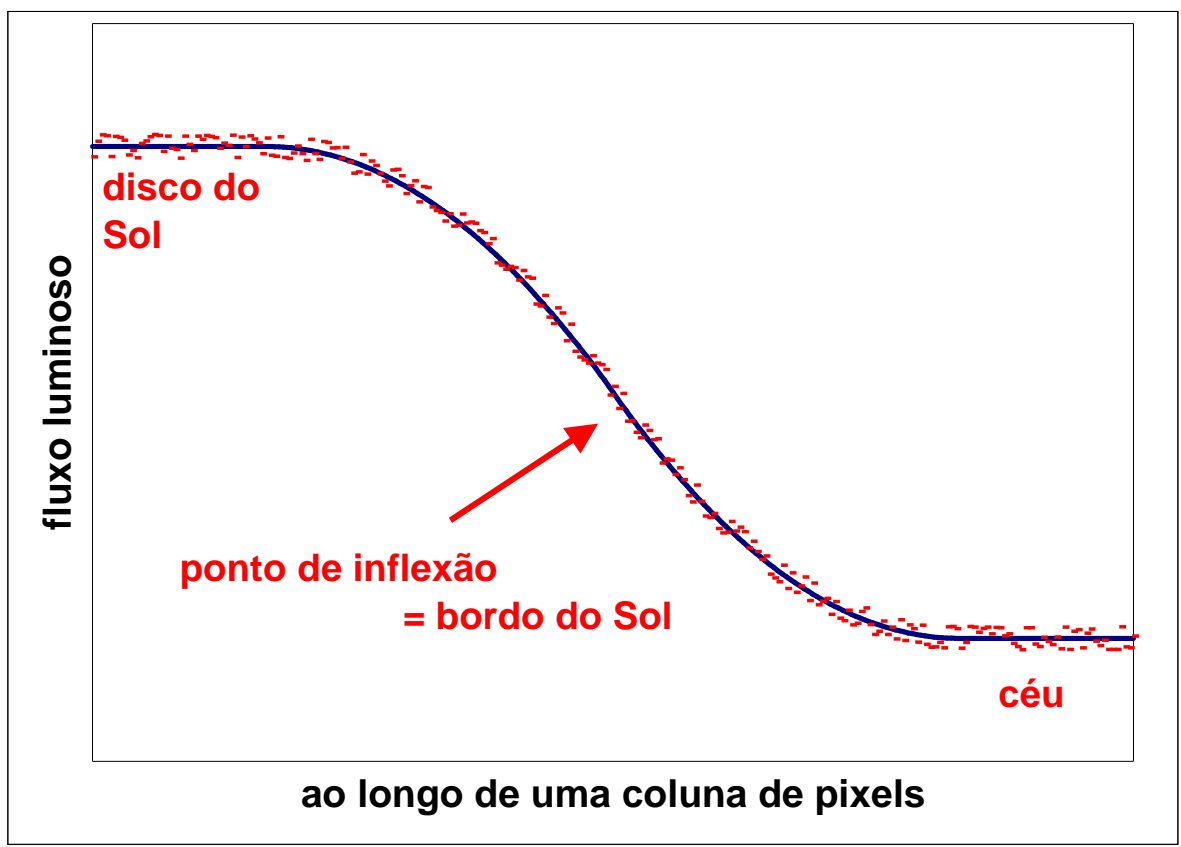

O bordo solar é constituído pelo conjunto destes 256 pontos. Ocorre, porém, que estes pontos geralmente formam uma curva ruidosa em função da ação da atmosfera. É preciso passar por eles uma curva mais definida que indique o bordo solar. Ajusta-se a estes pontos uma parábola e não um arco de círculo, isto porque a parábola minimiza defeitos óticos, responde melhor à forma retangular dos pixels, e contempla o movimento do disco solar durante os 20ms de integração [6]. Uma parábola para o bordo direto e uma parábola para o bordo refletido conforme é indicado no esquema apresentado na Figura IV.

**Figura IV - Esquema da determinação do bordo solar. Ajusta-se uma parábola aos pontos do bordo solar anteriormente determinados.**

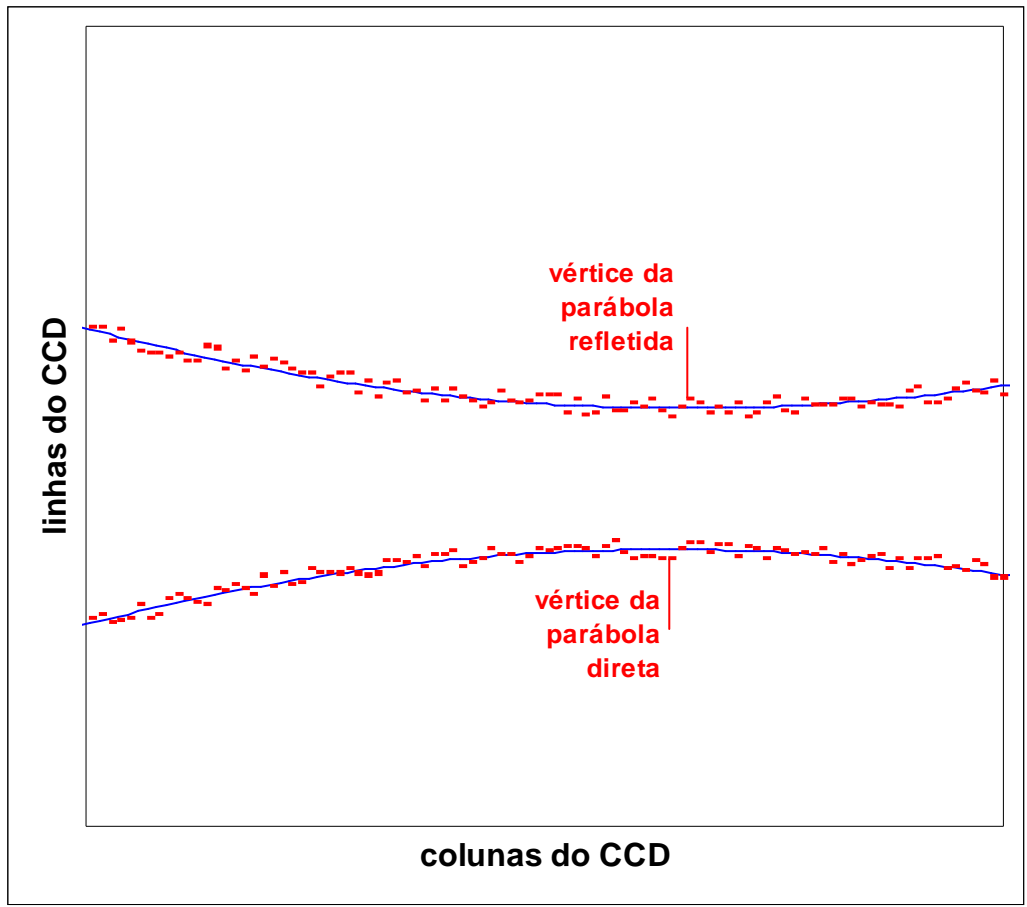

Para cada imagem obtida existem então duas parábolas. Tomam-se as posições dos vértices destas parábolas e o instante de tempo em que a imagem foi obtida. Estas posições em função do tempo definem duas curvas por onde se ajustam duas retas. Como uma das parábolas avança e a outra recua, estas retas têm inclinações opostas. O ponto de contacto destas retas define o instante de tempo em que os bordos direto e refletido do Sol se tocam. Este é o instante em que o bordo solar cruzou o almicantarado. Da mesma forma se encontra o instante em que o segundo bordo solar cruza o almicantarado. A Figura V mostra um esquema da determinação do instante de passagem do bordo solar pelo almicantarado.

**Figura V - Esquema da determinação do instante de passagem do bordo solar. O instante é determinado pelo ponto de contacto das retas ajustadas às posições sucessivas dos vértices das parábolas que determinam o bordo solar.**

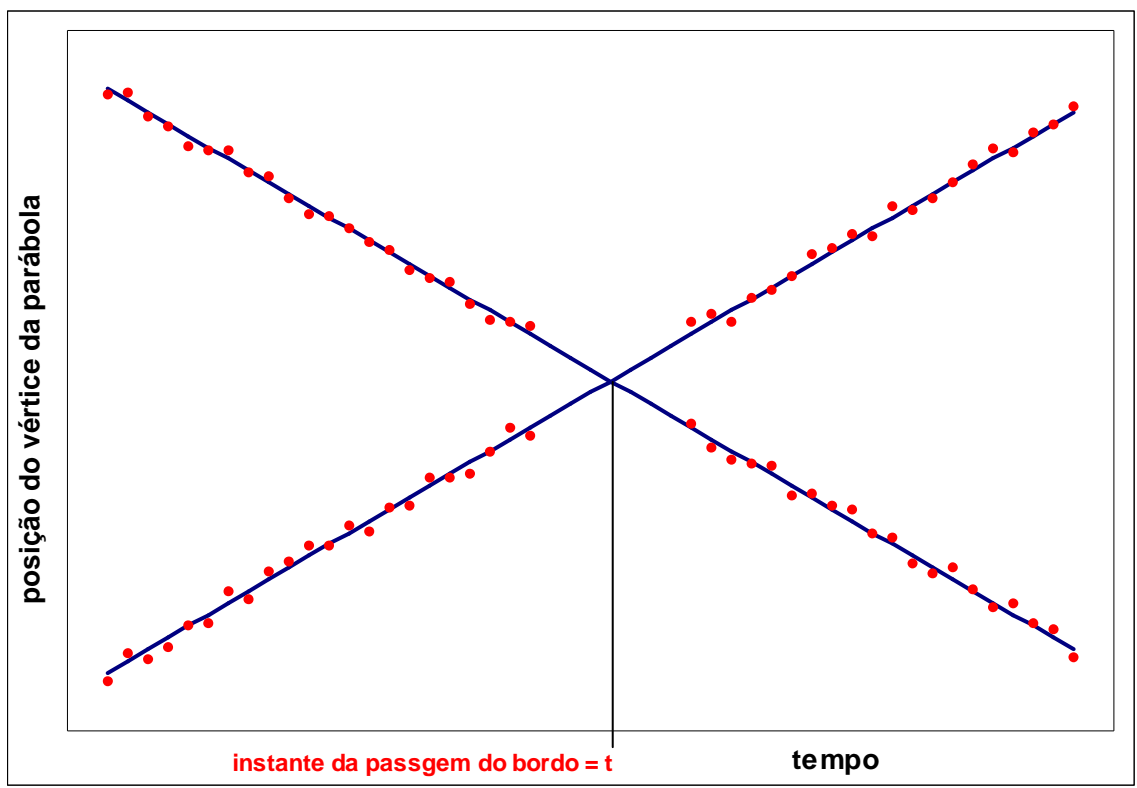

Um conjunto de programas desenvolvidos para tal, calcula, a partir dos instantes de passagem dos dois bordos pelo mesmo almicantarado, o diâmetro vertical observado do Sol. Após a aplicação das correções necessárias o diâmetro é reduzido para 1 UA [7].

Aos bordos do Sol ajusta-se, não apenas uma, mas, na verdade, três parábolas com critérios diferentes de desvio padrão para remoção de pontos observados. O critério se baseia num teste de quartís, com o fator multiplicativo variando entre 1,5 (mínima rejeição) e 3 (máxima rejeição). Se o número de pontos utilizados for inferior a 50, então nenhuma parábola é ajustada. Em 1997 foram definidos três níveis de critérios: 1,7, 2,0 e 2,5. Se as condições forem boas, três soluções são obtidas, caso contrário apenas as soluções para 1,7 e 2,0 alcançam resultado ou, mais raramente, apenas a primeira [8].

As imagens obtidas são analisadas pelos programas que calculam para cada observação os três valores do semidiâmetro solar e os respectivos erros de medida. Obtêm também a distância zenital e o azimute solar, a inclinação do sol, o fator de Fried, os instantes de passagem dos bordos solares, os erros de medida destes instantes, a largura em pixels do bordo direto e do bordo refletido, o ajuste da parábola direta e o da refletida, o desvio padrão dos pontos da parábola direta e, da refletida, e a decalagem. Para cada seção de observações têm-se também os instantes inicial e final, e as temperaturas do ar e do mercúrio, a pressão atmosférica e a umidade do ar nestes instantes.

O fator de Fried descreve a qualidade do 'seeing' da atmosfera. É definido como o comprimento de onda observado, dividido pela largura a meia altura de uma imagem pontual espalhada pela ação da atmosfera. Este fator é calculado a partir dos dados de observação [9].

A decalagem é o desvio das linhas e colunas do CCD no instante da passagem do bordo, causado por defeitos de orientação do prisma ou do CCD.

Diante da câmera do CCD, há dois filtros de luz que definem uma banda de freqüências que podem ser detectadas. O intervalo onde 50% da luz é transmitida vai de 523,0 nm até 691,0 nm. Sendo o máximo em 563,5 nm com índice de transmissão de 75% [10].

**OBSERVAÇÕES DURANTE O ANO DE 2001.**

Os dados do ano de 2001 têm especial interesse por duas razões. A primeira delas é que durante este ano ocorreu um segundo máximo da atividade solar em seu ciclo magnético. A análise pode revelar alguma relação entre variações do diâmetro solar e seu ciclo de atividade magnética. A segunda razão é a disponibilidade, pela primeira vez, de dados obtidos por outras equipes para o ano de 2001. Podemos então, comparar nossos dados com outros obtidos em outros locais e em outras condições. Dispomos de dados do CERGA na França e também de Antalya na Turquia.

As observações feitas no ON em 2001 se estendem do início de janeiro até quase ao final de setembro. A partir desta data o astrolábio foi temporariamente desativado para sofrer uma manutenção, voltando a operar somente nos últimos dias do ano. Assim, a série disponível de observações se estende do início de janeiro até o dia 21 de setembro. A distribuição de observações ao longo do ano é constante. Nota-se uma maior densidade de observações no início do ano, particularmente em janeiro e março com mais de 300 observações, e fevereiro e abril com mais de 200 observações. A partir de então há uma certa diminuição do número de observações. De maio até setembro há sempre um número entre 150 e 200 observações, com exceção de junho com pouco mais de 100 observações. Há ao todo 1890 observações retidas sendo 976 no lado leste, antes da passagem meridiana do Sol, e 914 a oeste, após a passagem.

Os pontos observados a leste são um pouco mais numerosos que os observados a oeste, particularmente de janeiro a abril. De maio a setembro há um maior equilíbrio de observações a cada lado. A Figura VI mostra a distribuição de pontos observados ao longo do ano.

**Figura VI - Distribuição das observações ao longo do ano.**

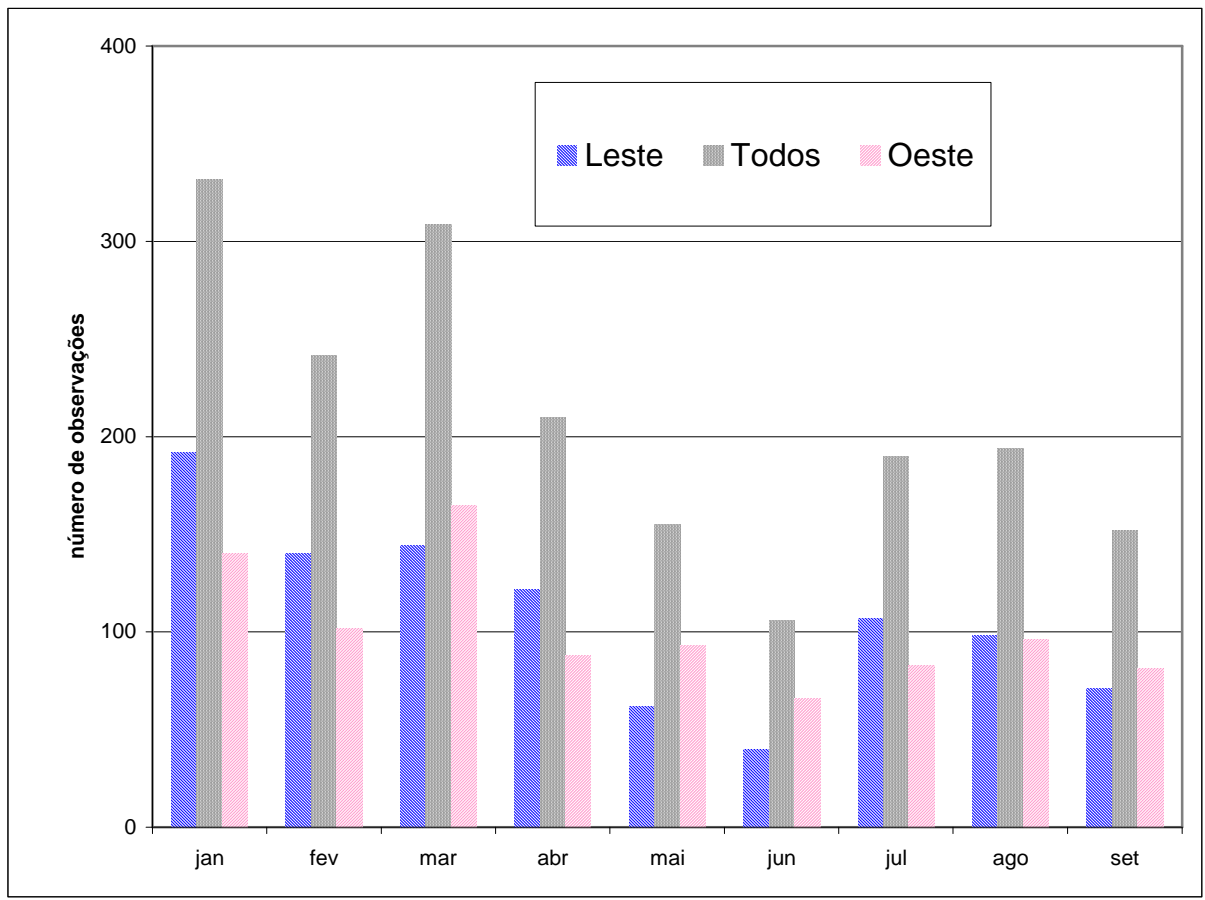

**PROCEDIMENTO INICIAL.**

Como já dissemos, o semidiâmetro solar calculado a partir das observações tem três séries de valores de acordo com o ajuste de uma parábola ao bordo solar. Há, portanto, três soluções que definem parábolas diferentes que se tocam em instantes ligeiramente diferentes, obtendo-se então três semidiâmetros diferentes para o Sol em cada observação realizada. Neste estudo consideramos apenas uma destas séries, que é aquela para a qual na construção das parábolas tomaram-se todos os pontos, exceto os que distam mais de 1,7 desvios-padrão de afastamento máximo da parábola traçada.

As três séries, segundo estudo anterior, apresentam comportamento semelhante, guardando entre si apenas uma pequena diferença para o valor do semidiâmetro do Sol. A série escolhida é a que retém um número maior de observações [11].

Os valores de semidiâmetro solar, assim obtidos, foram inicialmente submetidos a uma seleção na qual se descartaram observações mal sucedidas, comprometidas por algum motivo que levou a resultados extremados. Assim, foram descartados todos os registros nos quais o semidiâmetro solar ultrapassou 961,00 segundos de arco ou ficou aquém de 957,00 segundos de arco.

Foram também retirados os registros em que o erro de medida do semidiâmetro solar foi superior a um segundo de arco. Estes valores estão a mais de 25 vezes acima da sua média. Foram retirados ainda, os registros em que o erro na determinação do instante da passagem de um dos bordos foi superior a 0,1 segundo. Estes valores estão sempre a mais de 8 vezes acima da média dos valores.

Finalmente, foram retirados os registros das observações nas quais a inclinação do CCD foi superior a 2,5 graus. Neste caso, estes valores estão além de 4,7 vezes acima da média.

De acordo com estes critérios 194 registros foram removidos e descartados. Para a nossa análise foram selecionados os demais registros. Estes fazem um total de **1890**, sendo **976** de observações a leste e **914** de observações a oeste.

A média para o semidiâmetro solar dos valores aproveitados é **959,189** segundos de arco e o desvio padrão é de **0,610** segundos de arco. Para as observações a leste a média é **959,259** segundos de arco e o desvio padrão é de **0,643** segundos de arco. Para oeste a média é **959,115** segundos de arco e o desvio padrão **0,564** segundos de arco. Os gráficos das distribuições destes pontos podem ser vistos na Figura VII. A esta distribuição foi ajustada uma Gaussiana e o teste do Chi-quadrado forneceu um valor de 66,10233, sendo que para 20 graus de liberdade, como é o caso, um valor superior a 39,997 indica que os pontos seguem uma distribuição normal com 99,5% de certeza. Esta normalidade aparece por conta dos intervalos escolhidos, já que na realidade não há um valor médio, mas uma tendência em torno da média.

**Figura VII - Histograma dos valores observados de semidiâmetro do Sol, distribuídos em 20 intervalos de 0,2 segundos de arco, desde 957,0**

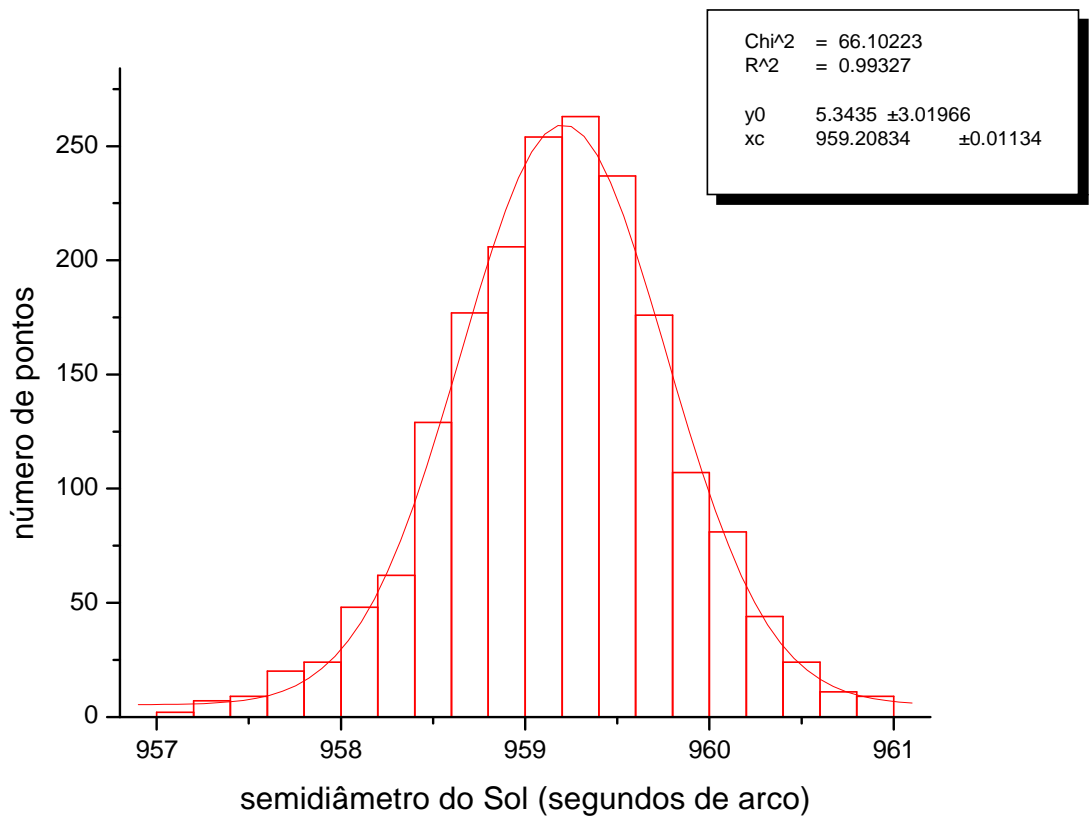

**até 961,0 segundos de arco e uma curva normal ajustada.**

Em todas as séries até hoje obtidas, os valores a oeste têm sempre um desvio padrão menor que os valores obtidos a leste. Isto deriva de condições mais estáveis para observação à tarde do que pela manhã.

Na Figura VIII pode-se ver o comportamento temporal do semidiâmetro solar obtido destas observações. Como os valores observados são muito dispersos, para se perceber sua evolução temporal, traçou-se sua média móvel. É uma média para 100 pontos.

**Figura VIII - Observações do semidiâmetro do Sol em duas séries: antes da passagem meridiana em azul e depois da passagem meridiana em**

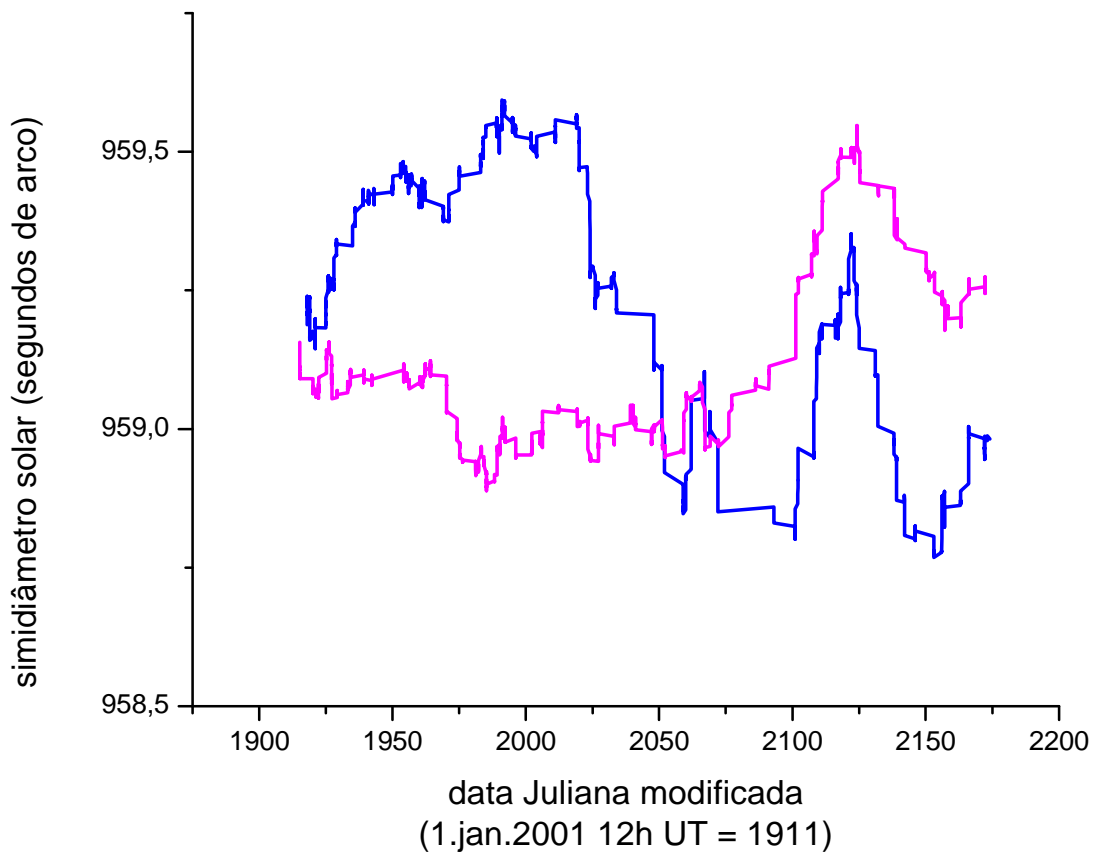

**magenta. As curvas representam as médias móveis a cada cem pontos.**

A escolha de 100 pontos para a média móvel leva a se tomar uma média em torno de um mês de dados uma vez que há pouco mais de 900 pontos para nove meses de observação em cada lado. O desvio padrão das observações do semidiâmetro solar está em torno de 0,6 segundos de arco. Assim, para uma média móvel de 100 pontos este desvio é de cerca de 0,06 segundos de arco, que é um valor cerca de 0,006% do valor de semidiâmetro solar. Desta forma, este valor é inferior às variações de fluxo do Sol que estão em torno de 0,01% e a nossa série, representada por uma média móvel de 100 pontos pode mostrar possíveis variações do semidiâmetro do Sol que seriam comensuráveis com as variações de seu fluxo de energia.

Os valores são mostrados em duas séries, uma série para as observações a leste e a outra para as observações a oeste. No eixo dos tempos foi colocada a data Juliana modificada, que vem a ser a data Juliana diminuída de 2.450.000 dias, desta forma os valores no eixo dos tempos variam entre um pouco mais que **1900** dias e um pouco menos que **2200** dias, ou seja, um período de quase 270 dias que corresponde aos meses de janeiro a setembro de 2001.

**Figura IX - Duração das observações do semidiâmetro solar, em função da data observada. Os valores são máximos próximos do solstício de inverno e mínimos próximos ao solstício de verão.**

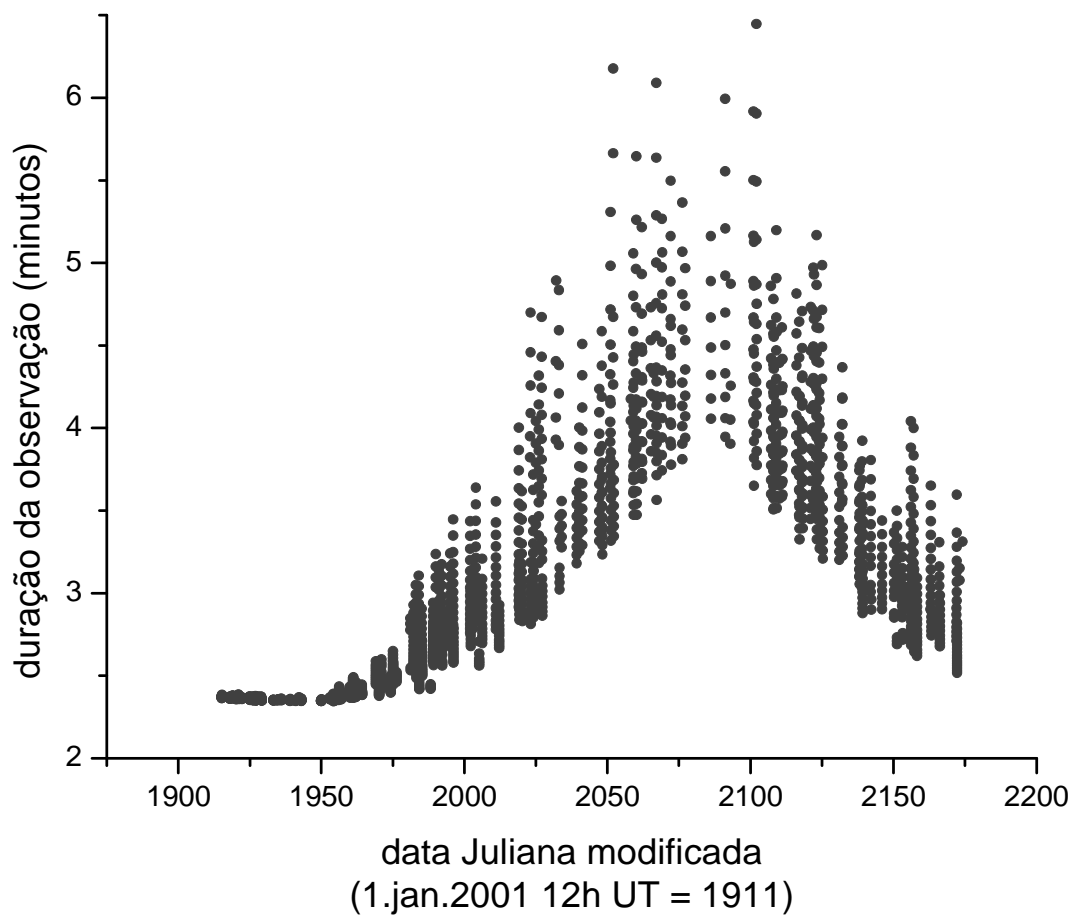

A Figura IX mostra que a duração das observações, isto é, o tempo que o Sol leva para cruzar a linha de altura zenital constante, é bastante variável, dependendo da época no ano e de sua posição no céu. Na série estudada este tempo foi de um mínimo de pouco mais de dois minutos a um máximo de quase sete minutos. Durante o verão, quando o Sol tem um caminho mais ascendente ou descendente, os tempos são menores e quase não há diferenças entre as várias observações do mesmo dia, mas, a medida em que se vai para o solstício de inverno, os tempos são maiores e maiores são também as diferenças entre as observações do mesmo dia. Isto se dá porque nesta época o Sol tende a fazer no céu arcos cada vez mais paralelos ao horizonte, subindo e descendo pouco no céu. Nesta figura podemos ver, no meio do ano, destacadas, cada uma das observações bem como cada grupo de observações no mesmo dia.

**Figura X - Instante da observação em Tempo Universal em função da data de observação. Os valores acima correspondem a observações a oeste e os valores abaixo corresponde a observações a leste.**

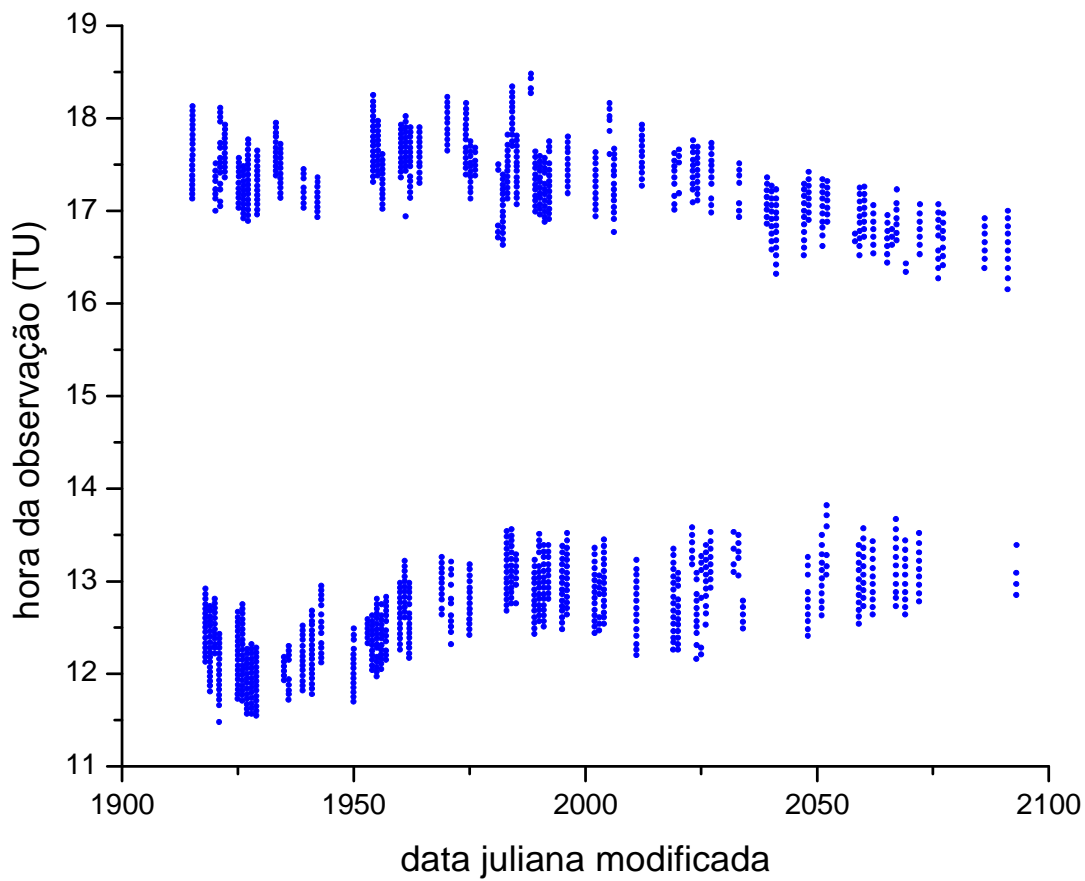

A Figura X mostra o horário de observação em função da evolução temporal. Pode-se ver que há dois grupos bem distintos de pontos que são as observações a leste em baixo e as observações a oeste em cima, ambos mais ou menos simétricos em relação à passagem meridiana do Sol – em torno de 15h UT. No início do ano, onde se configura o verão, os pontos são mais afastados da média enquanto que nas outras estações os pontos se aproximam da média, isto, porque a necessidade de se observar em baixas distâncias zenitais, obriga o observador a fazer seções mais próximas da passagem meridiana, quando não está no verão. O gráfico permite também perceber o prolongamento das seções de observação e revela também, pela ausência de pontos, os dias em que não foi possível observar.

Na Figura XI podemos ver o horário de observação em função dos desvios do azimute solar para o norte. Aqui também se vêem dois grupos bem distintos de pontos. Acima as observações a oeste, que são feitas após a passagem meridiana, e abaixo as observações a leste. Lembramos que não há observações próximas do azimute Norte, já que o astrolábio só permite observações para distâncias zenitais superiores a $25^{o}$, os menores desvios de azimute para a direção norte têm valores ligeiramente inferiores a $20^{o}$.

No início do ano, por ser verão, os desvios para o norte são maiores e estes pontos aparecem nas maiores abscissas. Os pontos das menores abscissas correspondem a observações mais próximas do inverno, onde as diferenças de azimute do Sol para a direção norte são menores. Pode-se ver as séries de observações na mesma seção, aumentando os desvios para o norte, na medida em que se afastam da passagem meridiana. Estes aumentos são bem pronunciados no inverno, onde a trajetória do Sol na esfera celeste tende a seguir mais os arcos paralelos ao horizonte, e pouco pronunciados no verão.

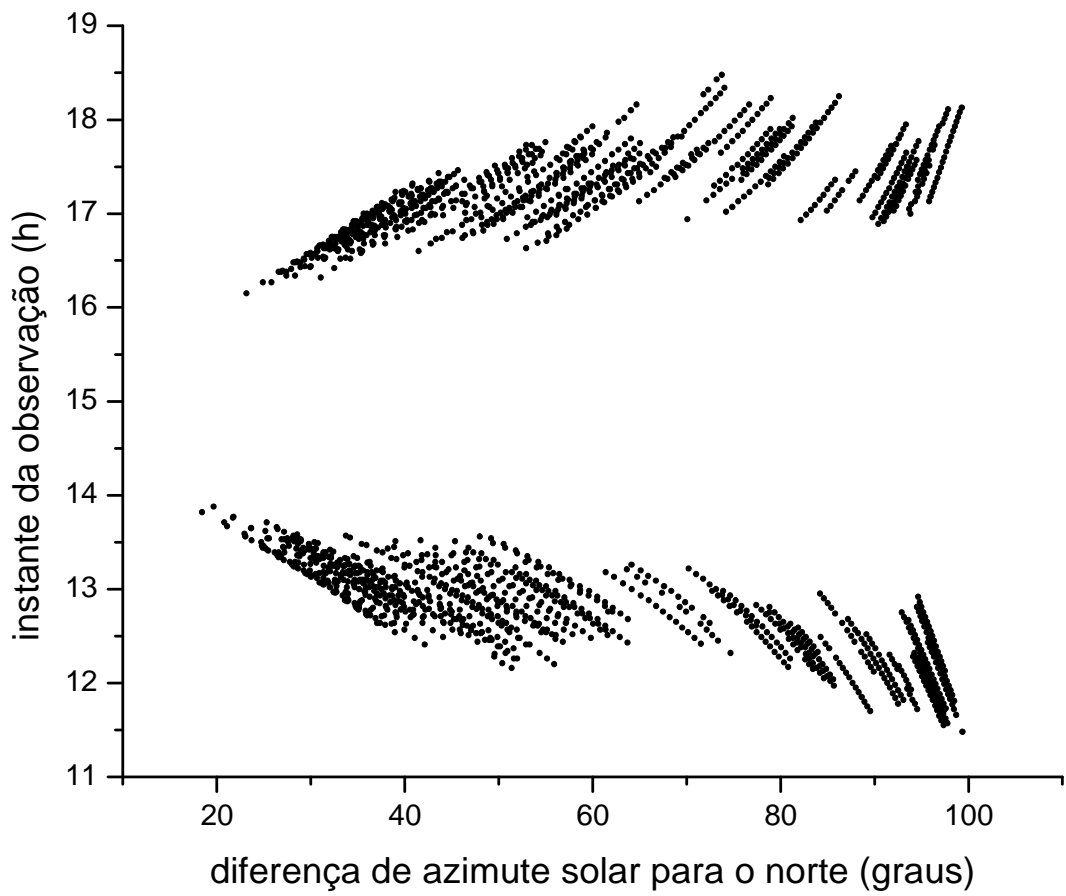

**Figura XI - Instante da observação em Tempo Universal em função da diferença de azimute do Sol para a direção Norte.**

# ESTABILIDADE DO PRISMA OBJETIVO.

Muito embora a qualidade metrológica, os valores obtidos de semidiâmetro solar podem conter uma variedade de erros decorrentes de inúmeros fatores, alguns erros são causados pela forma como se observa, outros são introduzidos pelo próprio instrumento, outros ainda pela atmosfera, além de outros fatores. A nosso favor temos uma grande quantidade de observações que oscilam aleatoriamente em torno do valor correto, ou melhor, em torno da sua tendência. A grande quantidade de valores observados pode ser útil também para se detectar um comportamento nos resultados que tenha sofrido a influência de determinado fator. Assim, uma inspeção adequada pode apontar um tipo de influência que tenha alterado de alguma maneira os valores medidos

Um astrolábio adaptado para observar o Sol pode observá-lo em qualquer distância zenital escolhida dentro de determinada faixa. Para que isto seja possível os espelhos do prisma objetivo devem ser deslocados de um certo ângulo até serem fixados em uma posição onde se possa observar um almicantarado desejado. No nosso caso, para se manter o ângulo, um conjunto de molas é utilizado. Se durante a observação, que dura poucos minutos, o conjunto de molas sofrer alguma alteração, haverá um erro de leitura que é tanto maior quanto maior for o tempo de observação. Entretanto este erro é diferente para os lados leste e oeste porque no primeiro caso o Sol está diminuindo sua distância zenital e no outro caso o Sol está aumentando a distância zenital, enquanto que as molas devem atuar num só sentido, de fechar ou de abrir o ângulo entre os prismas, fazendo com que a distância zenital observada seja respectivamente aumentada ou diminuída. Assim, para um dos lados haverá um acréscimo de tempo para o Sol cruzar a linha de almicantarado desejado enquanto que para o outro haverá um decréscimo. Esta diferença será de alguma forma proporcional ao tempo de observação.

Há evidências de que tal fato ocorre com o astrolábio do ON. Durante a observação da passagem do Sol pela altura zenital desejada as faces dos prismas tendem a se fechar, diminuindo o ângulo entre elas, fazendo com que a distância zenital observada, que deveria ser fixa, diminua ao longo da observação. Assim os valores de semidiâmetro do Sol a leste são ligeiramente aumentados e, os valores a oeste, diminuídos. Embora tal desvio seja bastante pequeno, ainda assim, pode ser detectado nas observações do semidiâmetro do Sol.

Para avaliar este desvio construímos gráficos das observações a leste e a oeste em função dos tempos de observação, a fim de determinar alguma tendência contrária em cada um dos gráficos. Infelizmente, este tipo de procedimento mostrou-se ineficaz, pois, a grande maioria dos pontos observados em cada um dos lados tem tempos curtos de observação, e poucos pontos têm tempos de observação mais longos. Os tempos de observação se estendem de **2,3** minutos a **6,9** minutos. Setenta e um por cento deles ficam entre 2,3 e 3,45 minutos, isto é, no primeiro quartil da divisão dos tempos. Noventa e cinco por cento deles ficam na primeira metade desta divisão, enquanto que apenas sete pontos observados o que significa 0,4% deles ficam no último quartil que são justamente os pontos mais afetados. Assim, as curvas ajustadas às funções dos semidiâmetros pelos tempos de observação não têm significado, pois são fortemente desviadas pelos poucos pontos nos extremos dos tempos maiores. E, desta forma, não há como se detectar a influência do tempo nas medidas de semidiâmetro. Ou seja, não há como encontrar a influência do tempo que se torna pequena diante dos grandes desvios provocados pelos pontos não interessantes.

A Figura XII mostra, para os pontos observados a leste, os semidiâmetros observados em função da duração das observações. Aos pontos foram ajustadas uma reta e uma parábola. Pode-se notar a grande concentração de pontos nos tempos menores. A Figura XIII mostra exatamente o mesmo para pontos observados do início do ano até 21 de abril. Apesar destes pontos corresponderem a 54% do total de pontos, os ajustes deles a uma reta e a uma parábola são completamente diferentes dos ajustes observados na figura anterior. Esta figura demonstra que os demais pontos, apenas 46% do total, porém, pontos com as durações mais longas, desviam em muito as tendências do total de pontos para se ajustarem a retas e parábolas. As Figuras XIV e XV mostram o mesmo efeito para os pontos observados a oeste que se concentram mais fortemente nos menores tempos para a série parcial que vai até 21 de abril.

Entretanto, se um gráfico dos semidiâmetros em função dos tempos de observação pode não ajudar, podemos recorrer a um gráfico dos semidiâmetros em função da data de observação. Neste gráfico aparece a mesma tendência dos valores de semidiâmetro em função dos tempos de observação uma vez que do início do ano até o final de junho os tempos de observação tendem a aumentar pois neste período estamos caminhando do solstício de verão para o solstício de inverno. Isto deve causar tendências opostas para as séries a leste e a oeste. De julho em diante as tendências devem se reverter, pois então, estaremos caminhando do solstício de inverno para o de verão e os tempos de observação passam a diminuir.

**Figura XII - Semidiâmetro observado a leste em função do tempo de duração da observação. Os pontos foram ajustados a uma reta e a uma parábola. A reta é mais alta que a parábola à esquerda e mais baixa à direita do gráfico. Note-se a grande concentração de pontos para os menores tempos.**

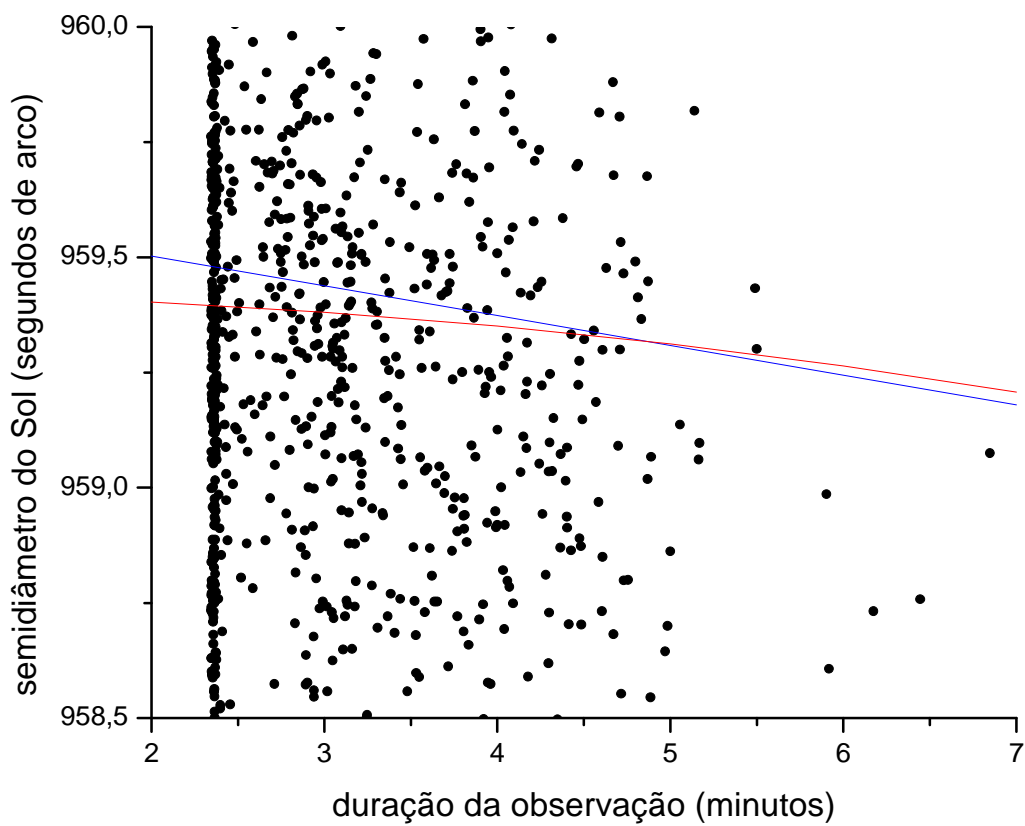

**Figura XIII -** Semidiâmetros observados a leste do início do ano até 21 de abril, em função do tempo de duração das observações. Os pontos foram ajustados a uma reta e a uma parábola. Correspondem a 54% de todos os pontos da série. Note-se a mudança completa nas tendências dos pontos.

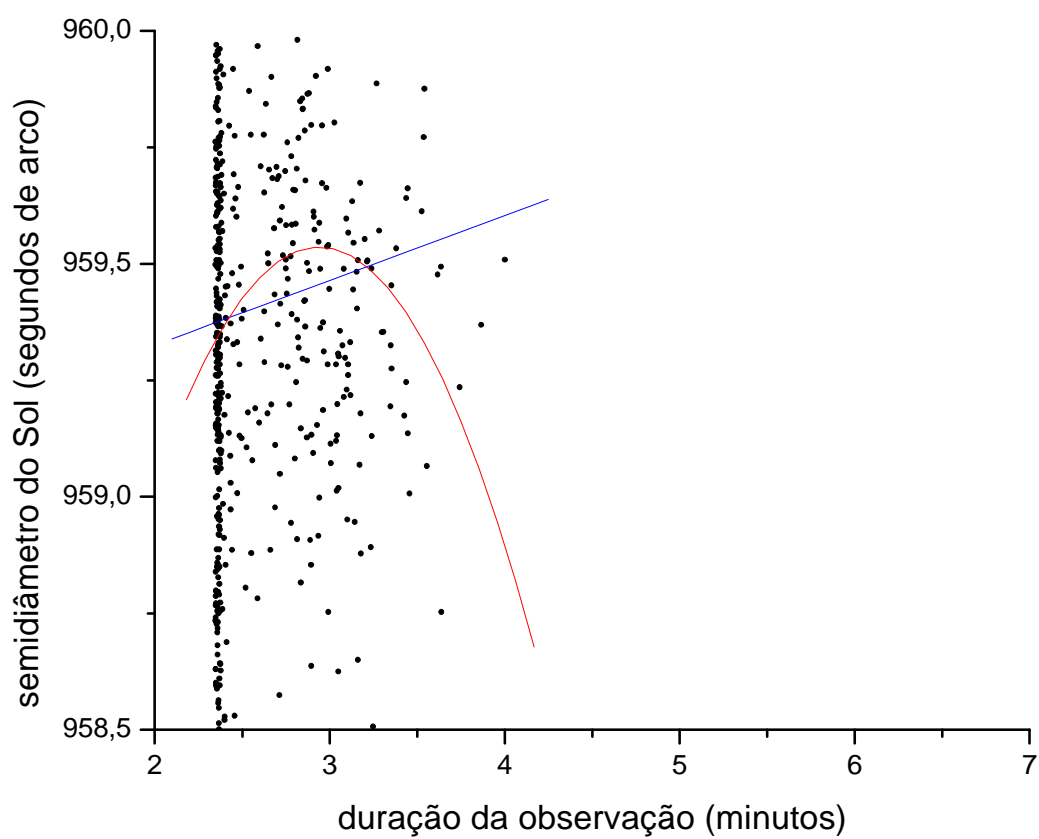

**Figura XIV - Semidiâmetro observado a oeste em função do tempo de duração da observação. Os pontos foram ajustados a uma reta e a uma parábola. A reta é mais alta que a parábola à esquerda e mais baixa à direita do gráfico.**

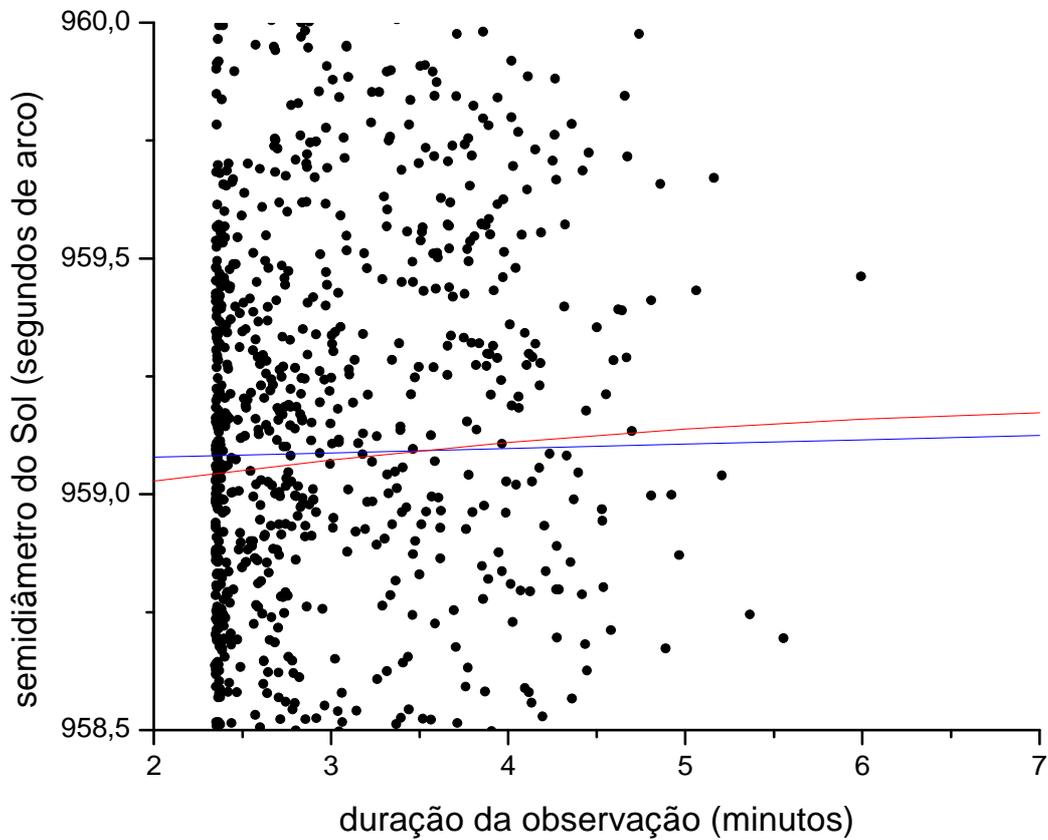

**Figura XV - Semidiâmetros observados a oeste do início do ano até 21 de abril, em função do tempo de duração das observações. Os pontos foram ajustados a uma reta e a uma parábola que se confundem. Os pontos se concentram muito nos menores tempos.**

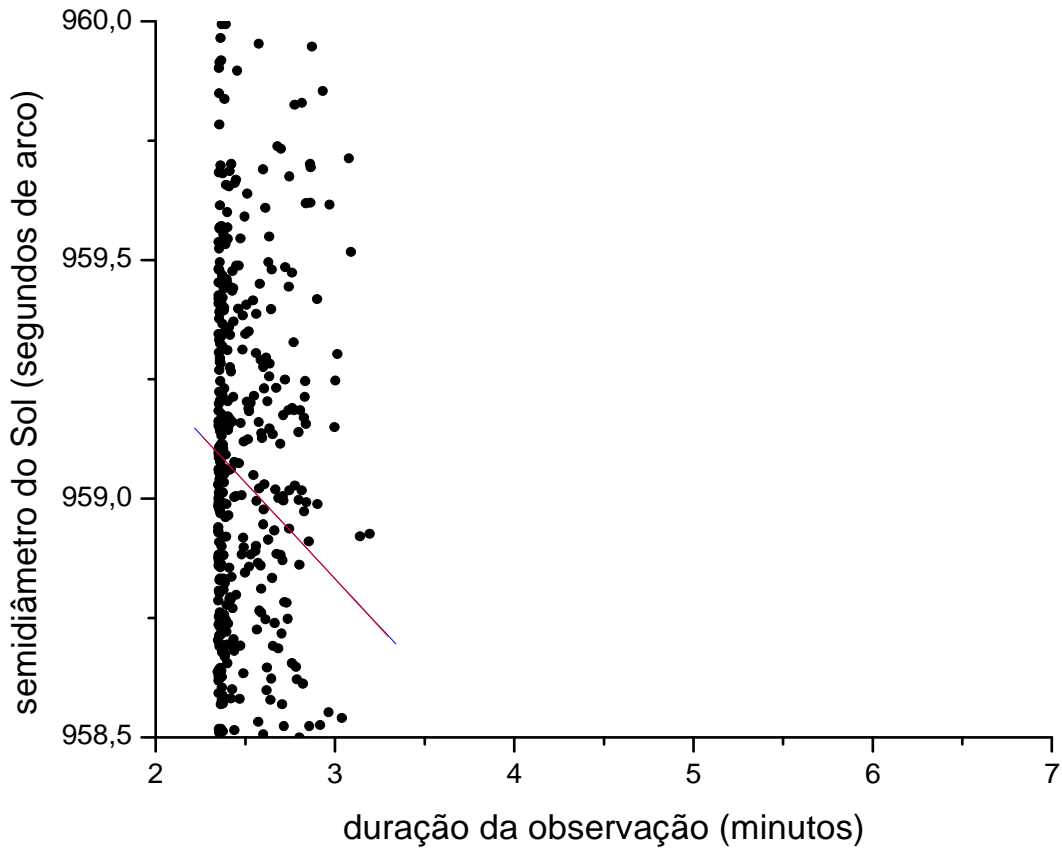

Ocorre aqui um outro problema, menos de 29% dos pontos observados ocorre depois do solstício de inverno e eles estão concentrados no início do período quando ainda são longos os tempos de observação, assim, estes pontos não são suficientes para reverter a tendência das séries.

Entretanto a tendência diferenciada para observações a leste a e a oeste pode, de fato, ser observada nos dados disponíveis. Ao se ajustar uma reta aos valores de semidiâmetro solar, observados a leste, em função da data Juliana da observação, obtém-se uma reta decrescente. Para valores observados a oeste, a reta é crescente. Ajustando-se parábolas aos pontos se observa também tendências diferentes, para os pontos a leste a concavidade da parábola é voltada para baixo, já para os pontos a oeste, esta é voltada para cima. As retas assim obtidas bem como as parábolas podem ser vistas na Figura XVI. Nesta figura também se vê uma reta ajustada às observações tomadas globalmente, isto é, somadas as observações a leste e a oeste.

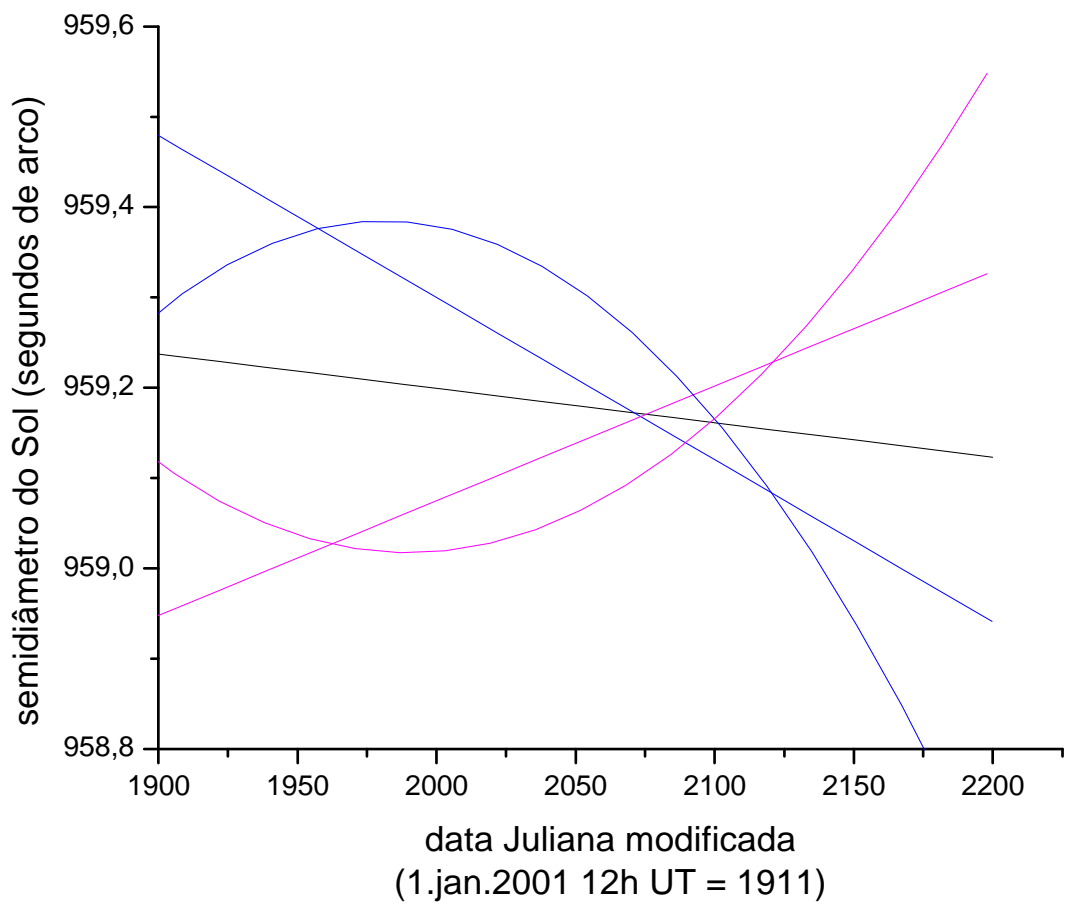

**Figura XVI - Séries de observações a leste em azul e a oeste em magenta, ajustadas linearmente e para parábolas. O total de pontos também está ajustado a uma reta.**

As retas ajustadas aos valores têm por função: **Y = A + B.X**, onde **Y** é o semidiâmetro solar e **X** é a data Juliana modificada. **A** é o coeficiente linear e **B** o coeficiente angular da reta. A Tabela III mostra estes coeficientes das retas ajustadas aos pontos a leste, a oeste e a todos os pontos globalmente.

**Tabela III – Coeficientes das retas ajustadas aos pontos observados.**

| Coeficiente | Valores a leste | Valores a oeste | Todos os valores |
|---|---|---|---|
| Linear (") | 962,891±0,508 | 956,534±0,473 | 959,961±0,358 |
| Angular ("/d) | -1,800.10⁻³±0,2511.10⁻³ | 1,270.10⁻³±0,2328.10⁻³ | -0,3808.10⁻³±0,1763.10⁻³ |

As parábolas ajustadas têm por função: **Y = A + B1.X + B2.X²**, onde **Y** é o semidiâmetro solar e **X** é a data Juliana modificada. **A** é o coeficiente independente, **B1** o coeficiente de primeiro grau e **B2** o coeficiente de segundo grau. A Tabela IV mostra estes coeficientes para as parábolas ajustadas aos pontos a leste e aos pontos a oeste.

**IV – Coeficientes das parábolas ajustadas aos pontos observados.**

| Coeficientes | Valores a leste | Valores a oeste |
|---|---|---|
| Independente (") | 898,803±16,358 | 1007,734±14,732 |
| Primeiro grau ("/d) | 0,06117±0,01606 | -0,04896±0,01445 |
| Segundo grau ("/d²) | -1,544.10⁻⁵±3,940.10⁻⁶ | 1,230.10⁻⁵±3,537.10⁻⁶ |

Ajustamos também polinômios de terceiro grau aos valores observados à leste e aos valores observados a oeste. Podemos ver na Figura XVII que estas curvas também exibem tendências opostas para cada um dos lados. Em cada uma destas duas curvas há um ponto onde a curvatura troca de lado, revertendo sua tendência. Este é o ponto de inflexão de cada curva. Nestas duas curvas este ponto ocorre exatamente no solstício de inverno o que indica que neste ponto há uma mudança de tendência nos pontos observados. Esta mudança de tendência ocorre na mesma data, tanto para os pontos a leste como para os pontos a oeste.

Os polinômios de terceiro grau ajustados têm por função: **Y = A + B1.X + B2.X² + B3.X³**, onde **Y** é o semidiâmetro solar e **X** é a data Juliana modificada. **A** é o coeficiente independente, **B1** o coeficiente de primeiro grau, **B2** é o coeficiente de segundo grau e **B3** o coeficiente de terceiro grau. A Tabela V mostra estes coeficientes da curva ajustada aos valores observados a leste e da curva ajustada aos valores observados a oeste.

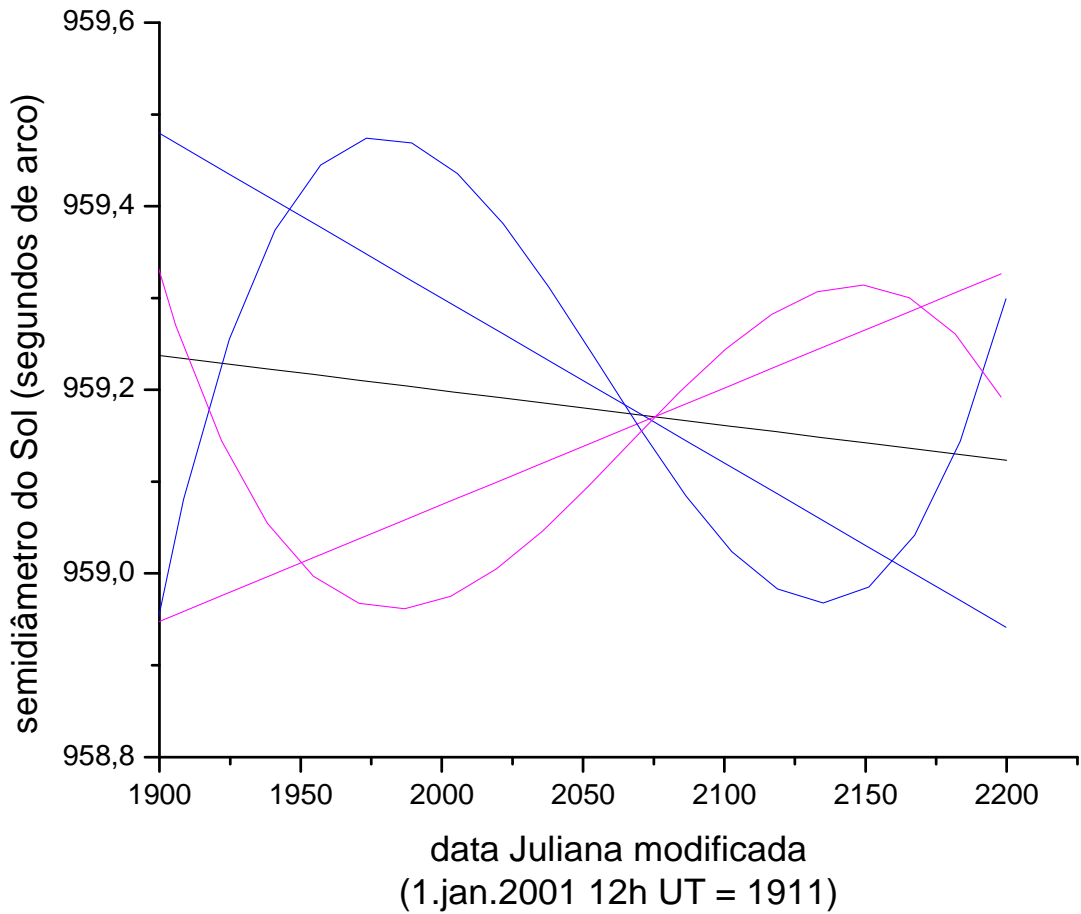

**Figura XVII -** Séries de observações a leste em azul e a oeste em magenta, ajustadas linearmente e para polinômios do terceiro grau. O total de pontos também está ajustado a uma reta.

**Tabela V – Coeficientes das curvas de terceiro grau ajustadas aos pontos observados.**

| Coeficientes | Valores a leste | Valores a oeste |
|---|---|---|
| Independente (") | -1337,409±496,853 | 2360,894±446,112 |
| Primeiro grau ("/d) | 3,360±0,7327 | -2,043±0,6571 |
| Segundo grau ("/$d^2$) | -1,640.$10^{-3}$±3,599.$10^{-4}$ | 0,9907.$10^{-3}$±3,224.$10^{-4}$ |
| Terceiro grau ("/$d^3$) | 2,651.$10^{-7}$±5,887.$10^{-8}$ | -1,599.$10^{-7}$±5,269.$10^{-8}$ |

O Sol observado a leste, ou observado a oeste não deve revelar tendências diferentes para a evolução de seu semidiâmetro. Esta diferença mostra há algum problema com o instrumento ou na forma de se observar. Não podemos afirmar que esta diferença se deva ao problema de estabilidade do prisma objetivo, ou exclusivamente a ele, mas, seja qual for a causa deste erro devemos retirá-lo da série. Para tal, fazemos com que tanto a tendência a leste como a tendência a oeste sejam corrigidas para a tendência média. Pelas tendências opostas dos dados a leste e a oeste, ajustados a polinômios de terceiro grau e, pela característica destas curvas mostrarem tendências opostas de curvatura, antes e após o solstício de inverno, a correção dos dados retirando-lhes suas tendências a se ajustarem a polinômios de terceiro grau seria uma boa opção

Antes, porém, de procedermos à correção, verificamos a consistência dos ajustes dos pontos observados aos polinômios de terceiro grau. Para tal ajustamos os pontos observados a um polinômio de terceiro grau executado com a utilização do método dos mínimos quadrados. A função ajustada tem a forma: **Y = A + B1.X + B2.X$^2$ + B3.X$^3$**, onde **Y** é o semidiâmetro solar e **X** é a data Juliana modificada. **A** é o coeficiente independente, **B1** o coeficiente de primeiro grau, **B2** é o coeficiente de segundo grau e **B3** o coeficiente de terceiro grau. A Tabela VI mostra os coeficientes da curva ajustada aos pontos a leste e da curva ajustada aos pontos a oeste.

**Tabela VI – Coeficientes de um polinômio de terceiro grau obtido por mínimos quadrados e ajustado aos pontos observados.**

| Coeficientes | Valores a leste | Valores a oeste |
|---|---|---|
| Independente (") | -1666,239±484,366 | 2665,819±434,010 |
| Primeiro grau ("/d) | 3,846±0,7143 | -2,495±0,6393 |
| Segundo grau ("/d$^2$) | -1,900.10$^{-3}$±4,000.10$^{-4}$ | 1,200.10$^{-3}$±3,000.10$^{-4}$ |
| Terceiro grau ("/d$^3$) | 3,000.10$^{-7}$±5,700.10$^{-8}$ | -2,000.10$^{-7}$±5,100.10$^{-8}$ |

Utilizamos então o método de Monte Carlo, sorteando pontos da série a leste e da série a oeste. Pela geração de números aleatórios, a cada um dos pontos da série foi dada uma chance de 30% de ser selecionado. Aos pontos selecionados ajustou-se um polinômio de terceiro grau. Desta forma tomamos 100 séries de pontos a leste e 100 séries de pontos a oeste e seus respectivos ajustes a polinômios de terceiro grau. Calculamos a média de cada um dos parâmetros dos polinômios ajustados obtendo os valores mostrados na Tabela VII, mais uma vez **A** é o coeficiente independente, **B1** o coeficiente de primeiro grau, **B2** é o coeficiente de segundo grau e **B3** o coeficiente de terceiro grau.

**Tabela VII – Médias dos coeficientes de polinômios de terceiro grau ajustados a 100 grupos de pontos observados sorteados.**

| Coeficientes | Valores a leste | Valores a oeste |
|---|---|---|
| Independente (") | -1700,00±880,00 | 2700,00±630,00 |
| Primeiro grau ("/d) | 4,000±1,300 | -2,500±0,9300 |
| Segundo grau ("/$d^2$) | $-1,900.10^{-3} \pm 6,400.10^{-4}$ | $1,200.10^{-3} \pm 4,600.10^{-4}$ |
| Terceiro grau ("/$d^3$) | $3,100.10^{-7} \pm 1,100.10^{-7}$ | $-1,900.10^{-7} \pm 7,400.10^{-8}$ |

As diferenças entre as médias dos parâmetros assim obtidos e os parâmetros obtidos pelo método dos mínimos quadrados divididas pelos erros deste último método, revelam valores sempre menores que 25%. Para os termos independentes temos 7% a leste e 8% a oeste, para os termos de primeiro grau 22% a leste e 1% a oeste, para os parâmetros de segundo grau não há diferenças dentro da precisão utilizada e para os termos de terceiro grau temos 18% a leste e 20% a oeste. Comparando a média de erro de cada parâmetro dada pelo método de Monte Carlo e o erro de cada parâmetro dado pelo método dos mínimos quadrados os valores estão sempre entre 1,45 e 1,93. Estes números mostram que os dois ajustes são bastante próximos. Assim, podemos afirmar que os pontos da série a leste e da série a oeste se ajustam no seu total ou em partes sorteadas de forma consistente a um polinômio de terceiro grau.

Podemos demonstrar que a correção linear dos dados observados é bastante próxima da correção pelo polinômio de terceiro grau. Para as curvas a leste, o desvio médio entre elas é de +0,01 segundos de arco com um desvio padrão de 0,17 segundos de arco. Para as curvas a oeste o desvio médio é de –0,02 segundos de arco e o desvio padrão de 0,13 segundos de arco. Considerando apenas o trecho onde as curvas tem contacto e onde tanto uma como outra curva têm a melhor representatividade, estes números caem ainda mais: para as curvas a leste o desvio médio é de –0,02 segundos de arco e o desvio padrão de 0,09 segundos de arco e para as curvas a oeste o desvio médio é de +0,02 segundos de arco e o desvio padrão de 0,06 segundos de arco. Lembrando que o desvio padrão dos pontos de semidiâmetro solar observados é de 0,61 segundos de arco, pode-se considerar que o ajuste linear e o ajuste ao polinômio de terceiro grau são muito próximos.

Dada a proximidade destes dois ajustes e pela simplicidade de se fazer uma correção levando-se em conta os ajuste de primeiro grau, optamos então por corrigir os dados tendo por base os ajustes ao primeiro grau.

Para executar a correção das tendências lineares opostas dos pontos a leste e dos pontos a oeste, fazemos com que tanto pontos a leste como pontos a oeste se ajustem à reta mediana, ou melhor, à reta que se ajusta aos pontos tomados em sua totalidade, isto é, pontos a leste mais pontos a oeste. A reta que se ajusta aos pontos a leste e a reta mediana se encontram no ponto onde o semidiâmetro solar **Y = 959,17** segundos de arco e a data Juliana modificada **X = 2071,0** dias. A reta que se ajusta aos pontos a oeste e a reta mediana se encontram em **Y = 959,17** segundos de arco e **X = 2077,2** dias.

Assim, para corrigir a tendência dos pontos a leste de modo que passem a ter a tendência média, soma-se aos pontos atuais: **(-0,38083+1,7959).10$^{-3}$ ΔX = 1,41477.10$^{-3}$ ΔX** onde **ΔX** é a diferença: **ΔX = (X-2071,0)**, isto é **X** menos o valor da data Juliana modificada onde as retas a leste e mediana se encontram. O valor do parênteses que multiplica **ΔX** é a diferença entre os parâmetros angulares da reta mediana e da reta a leste.

Para se corrigir a tendência dos pontos a oeste soma-se aos pontos originais: **(-0,38083-1,2704).10$^{-3}$ ΔX = -1,65123.10$^{-3}$ ΔX** onde **ΔX** é a diferença: **ΔX = (X-2077,2)**, isto é, **X** menos o valor da data Juliana modificada onde as retas a oeste e mediana se encontram. O valor do parênteses que multiplica **ΔX** é a diferença entre os parâmetros angulares da reta mediana e da reta a oeste.

As correções assim aplicadas foram de um mínimo de **–0,216** a um máximo de **0,267** segundos de arco, sua média, naturalmente, muito próxima de zero e seu desvio padrão de **0,141** segundos de arco. A Figura XVIII mostra o histograma destas correções. Temos uma distribuição quase uniforme das correções, ligeiramente diminuída no centro e nos valores máximos extremos. Esta distribuição está de acordo com o esperado já que a correção aplicada é determinada por duas retas que se afastam da reta média proporcionalmente às diferenças de datas.

**Figura XVIII - Histograma das correções aplicadas aos valores observados de semidiâmetro do Sol para corrigir erros introduzidos pela instabilidade da mola dos prismas da objetiva.**

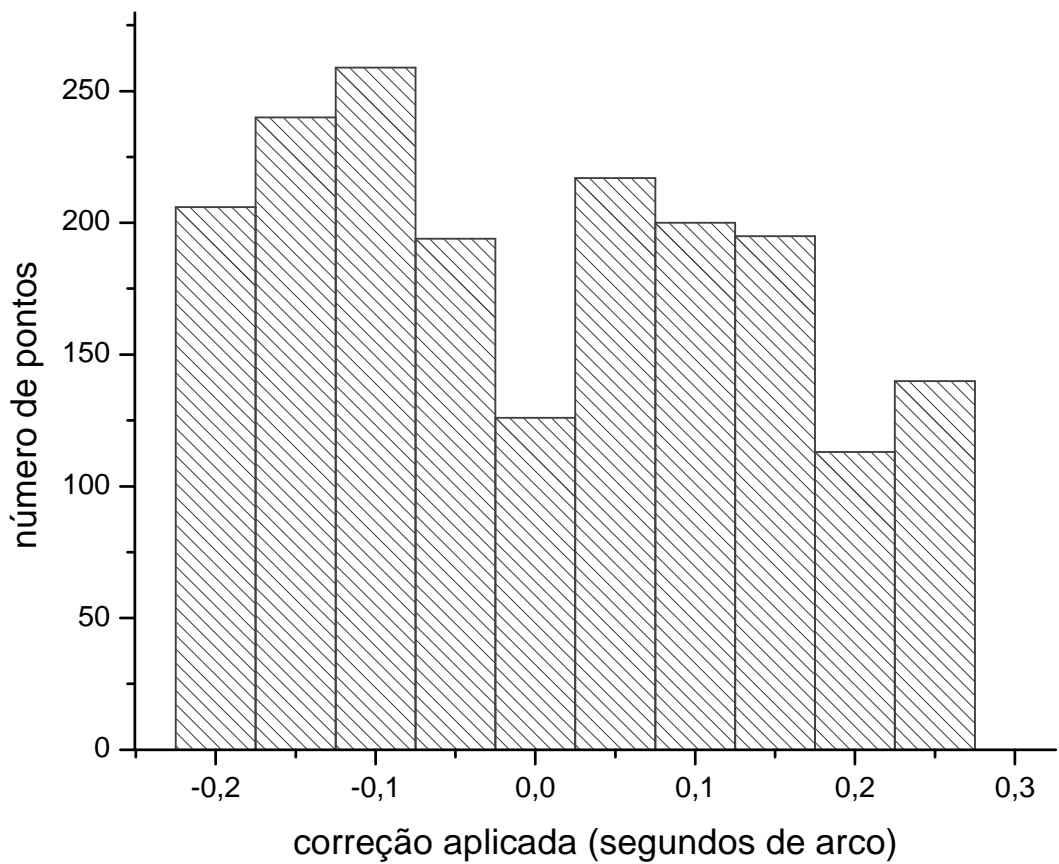

Uma vez aplicadas estas correções aos valores observados de semidiâmetro solar, a média para os valores a leste mudou então de **959,259** para **959,191** segundos de arco, e o desvio padrão caiu de **0,643** para **0,628** segundos de arco. Para os valores a oeste a média se deslocou de **959,115** para **959,190** segundos de arco e o desvio padrão caiu de **0,564** para **0,556** segundos de arco. A média do total de pontos foi de **959,189** para **959,191** segundos de arco e o desvio padrão caiu de **0,610** para **0,594** segundos de arco. O que se nota com esta correção é que os pontos a leste e a oeste foram corrigidos em média para valores mais próximos de sua média geral e os desvios padrão foram reduzidos em todos os casos. A Figura XIX mostra a distribuição dos pontos corrigidos. A esta distribuição está ajustada uma Gaussiana e o teste do Chi-quadrado forneceu um valor de 66,95766, sendo que para 21 graus de liberdade, como é o caso, um valor superior a 41,401 indica que os pontos seguem uma distribuição normal com 99,5% de certeza. Esta normalidade aparece por conta dos intervalos escolhidos, já que na realidade não há um valor médio, mas uma tendência em torno da média.

**Figura XIX - Histograma das observações do semidiâmetro solar depois de corrigidas da influência da instabilidade do prisma objetivo. Estão distribuídos em 21 intervalos de 0,2 segundos de arco, desde 957,0 até 961,2 segundos de arco e uma curva normal ajustada.**

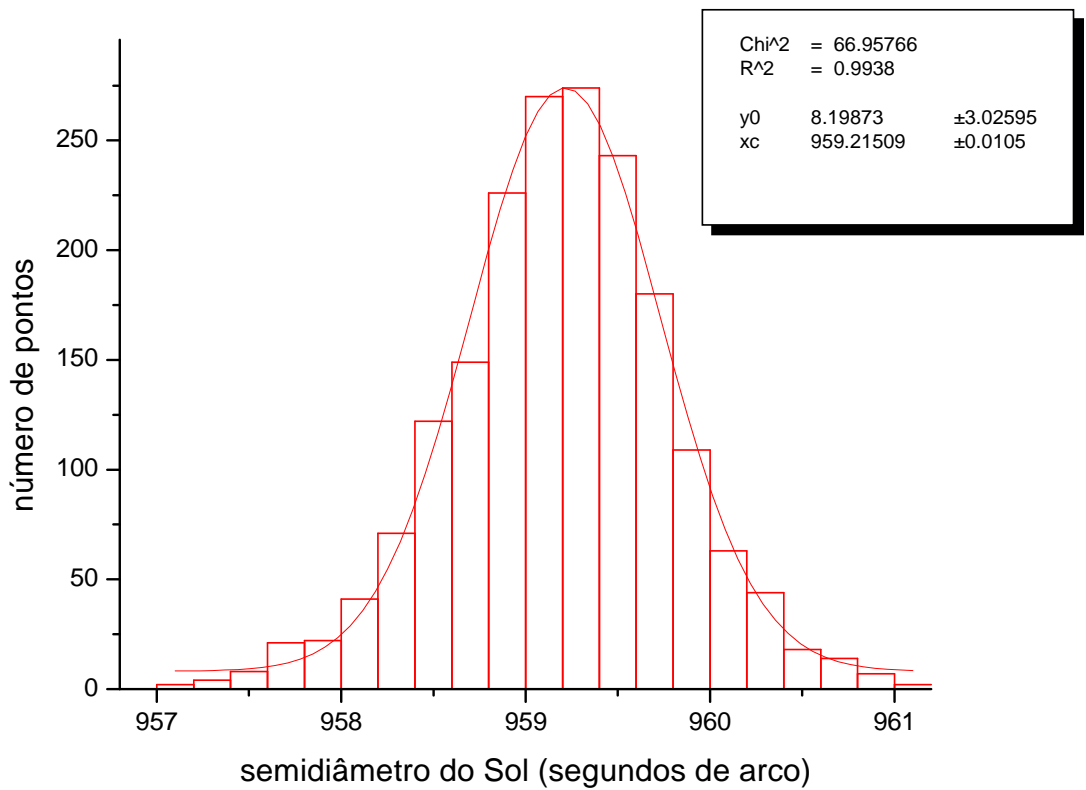

## PARÂMETROS DAS OBSERVAÇÕES.

Para cada observação do semidiâmetro solar temos disponível uma série de dados adicionais que podem ser chamados de parâmetros da observação. Cada um destes parâmetros pode separadamente ou em conjunto com outros, influir e alterar os valores medidos. Selecionamos vários destes parâmetros e verificamos separadamente e em conjunto, de que forma estes parâmetros poderiam estar influindo nos resultados. Foram selecionados, a data Juliana, a distância zenital do Sol, o azimute de observação do Sol, o parâmetro de Fried, o desvio padrão do ajuste da parábola ao bordo direto do Sol, o desvio padrão do ajuste da parábola ao bordo refletido do Sol, a temperatura do ar no instante médio da observação, a variação da temperatura durante a observação e a pressão atmosférica no instante médio da observação.

A possível influência destes parâmetros foi verificada por meio de ajustes lineares dos pontos observados aos parâmetros propostos e verificando-se o valor do coeficiente angular da relação linear bem como o desvio padrão do ajuste. Os pontos foram assim verificados, no seu conjunto total, nos subconjuntos a leste e a oeste e num conjunto total onde se tomaram os pontos a leste e os pontos a oeste com sinais invertidos, procurando-se assim estabelecer a existência de simetria nas influências de observações em lados opostos. O ajuste de retas aos dados foi executado com a utilização da técnica de mínimos quadrados.

Para se fazer estes cálculos, todos os parâmetros foram inicialmente normalizados, isto é, diminuídos de sua média e divididos por seu desvio padrão. Os cálculos assim realizados mostraram que quatro dos parâmetros inicialmente selecionados, tinham alguma influência nos dados de semidiâmetro do Sol, uma vez que os coeficientes angulares das retas ajustadas tinham valores relevantes e os desvios padrão eram, no mínimo, um terço menores que os valores. Esta análise nos conduziu a selecionar os quatro parâmetros que são: o azimute de observação do Sol, o fator de Fried, a temperatura média do ar durante a observação e o desvio padrão do ajuste da parábola ao bordo refletido do Sol. A influência causada pelo azimute de observação deve estar relacionada ao imperfeito nivelamento do instrumento observador. Resolvemos proceder a um estudo mais diretamente dirigido à influência deste parâmetro. Quanto aos outros três parâmetros os cálculos de suas influências lineares sobre os valores medidos apontam para as correções que são mostradas na Tabela VIII

**Tabela VIII - Correções devidas aos parâmetros analisados.**
**(valores em segundos de arco – parâmetros normalizados)**

| Parâmetro | Fator de Fried | Desvio padrão do ajuste | Temperatura do ar |
|---|---|---|---|
| Correção | $-0,7083.10^{-1}$ | $-0,2129.10^{-1}$ | $0,2368.10^{-2}$ |

A correção se aplica sobre o parâmetro normalizado. Os valores de semidiâmetro corrigidos são iguais aos valores anteriores somados à correção. A correção é a aplicação com o sinal invertido, do coeficiente angular da reta que se ajustou aos dados do semidiâmetro solar tomados em função do parâmetro analisado e multiplicada pelo próprio parâmetro normalizado. Assim procedendo, estamos retirando dos pontos a influência introduzida a eles pelo parâmetro em questão. Os valores normalizados são os valores diminuídos de sua média, e, divididos pelo desvio padrão. As três correções, assim calculadas, foram somadas para a obtenção da correção total do semidiâmetro observado.

Vale também acrescentar que todos os cálculos feitos para se obter estas correções tiveram por base de dados os valores observados do semidiâmetro solar já corrigidos dos efeitos de instabilidade do prisma objetivo que foram relatados no item anterior.

Das três correções a mais significativa deve-se ao fator de Fried em média três vezes maior que a correção devida ao desvio padrão do ajuste à parábola refletida, esta ainda cerca de dez vezes mais significativa que a correção devida à temperatura do ar durante a observação. As correções totais aplicadas foram de um mínimo de **–0,238** a um máximo de **0,255** segundos de arco sendo a média, naturalmente, muito próxima de zero e o seu desvio padrão igual a **0,067** segundos de arco. A Figura XX mostra um histograma da distribuição das correções. Esta distribuição é bem densa no centro, onde as correções são pequenas caindo rapidamente para os extremos onde ocorrem as correções maiores.

Uma vez aplicadas estas correções aos valores observados de semidiâmetro solar, a média para os valores a leste mudou de **959,191** para **959,186** segundos de arco, o desvio padrão caiu de **0,628** para **0,621** segundos de arco. Nos valores a oeste a média se deslocou de **959,190** para **959,195** segundos de arco e o desvio padrão de **0,556** para **0,558** segundos de arco. A média do total de pontos permaneceu igual a **959,191** segundos de arco e o desvio padrão caiu de **0,594** para **0,591** segundos de arco. Nota-se com esta correção que os pontos a leste e a oeste foram corrigidos para valores opostos em relação à média geral, se afastando dela. O desvio padrão foi ligeiramente reduzido para os pontos a leste, mas aumentou, ainda menos, para os pontos a oeste, de modo que, para o conjunto completo de pontos, o desvio padrão também foi ligeiramente reduzido. A Figura XXI mostra a distribuição dos pontos assim corrigidos. A esta distribuição está ajustada uma Gaussiana e o teste do Chi-quadrado forneceu um valor de 80,7964, sendo que para 21 graus de liberdade, como é o caso, um valor superior a 41,401 indica que os pontos seguem uma distribuição normal com 99,5% de certeza. Esta normalidade aparece por conta dos intervalos escolhidos, já que na realidade não há um valor médio, mas uma tendência em torno da média.

**Figura XX - Histograma das correções aplicadas aos valores de semidiâmetro do Sol para corrigir erros introduzidos pelos parâmetros: fator de Fried, desvio padrão do ajuste da parábola refletida e temperatura do ar.**

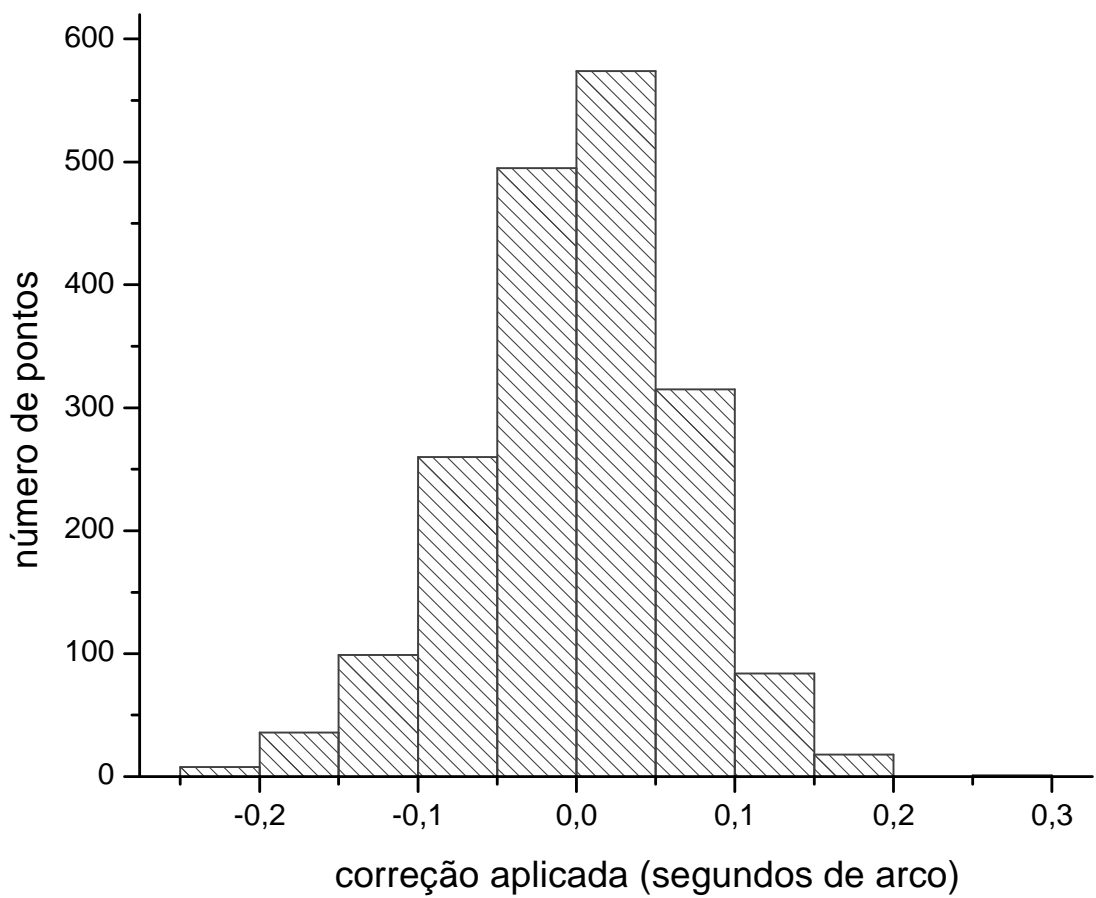

**Figura XXI - Histograma das observações do semidiâmetro solar depois de corrigidas da influência de parâmetros de observação. Estão distribuídos em 21 intervalos de 0,2 segundos de arco, desde 957,0 até 961,2 segundos de arco e uma curva normal ajustada.**

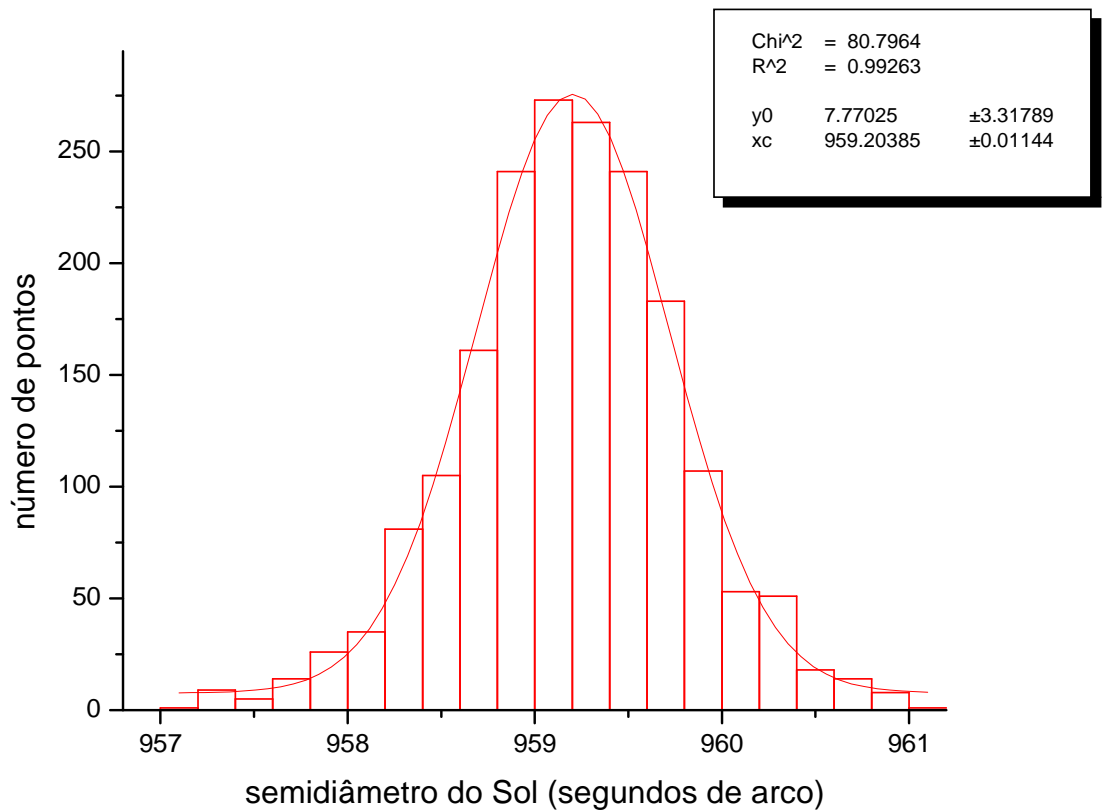

# O AZIMUTE DE OBSERVAÇÃO.

Um defeito de nivelamento do aparelho observador tem influência na medida do semidiâmetro solar, que pode ser medido em função do azimute de observação do Sol. Um pequeno desvio da vertical do eixo em torno do qual o aparelho gira no seu apontamento azimutal pode causar erros de observação que dependerão do ângulo de apontamento.

Os ângulos de azimute são medidos a partir da direção Sul e no sentido sul, oeste, norte, leste e retornando ao sul. Assim, azimute zero aponta para o Sul; azimute 90$^o$ aponta para Oeste, azimute 180$^o$ aponta para Norte e azimute 270$^o$ aponta para Leste. Desta forma, quando o Sol está a leste, na latitude do Rio de Janeiro, seu azimute pode variar entre algo um pouco mais que 180$^o$ até algo um pouco mais que 270$^o$, no nosso caso, mais precisamente estes limites ficaram entre 198,46$^o$ e 279,37$^o$. Quando o Sol está a oeste este ângulo pode variar de algo pouco menor que 90$^o$ até pouco menos de 180$^o$, e no nosso caso, os limites foram 80,71$^o$ e 156,82$^o$.

Ao procurar alguma influência deste ângulo na medida do semidiâmetro do Sol, não podemos considerar, pura e simplesmente estes valores de ângulo, que não têm nenhum significado, mas sim, o desvio deste ângulo para uma direção de simetria, por exemplo, ao norte. Assim, procuramos verificar de que forma se comportam os valores medidos em função do valor absoluto dos desvios de azimute de observação solar para a direção Norte.

Os pontos observados, quando em função dos desvios do azimute de observação do Sol para a direção Norte, se ajustam a uma reta cuja equação é: **Y = A + B.X**, onde **Y** é o semidiâmetro solar e **X** é o valor absoluto do desvio do azimute de observação do Sol para a direção Norte. A é o coeficiente linear da reta e B o coeficiente angular. A Tabela IX mostra estes coeficientes para as retas ajustadas aos pontos a leste, a oeste e a todos os pontos globalmente.

Na reta que se ajusta aos pontos a leste o valor médio de semidiâmetro de **959,181** segundos de arco corresponde a um desvio de ângulo de **58 $^o$,590**. Na reta que se ajusta aos pontos a oeste o valor médio de semidiâmetro de **959,195** segundos de arco corresponde a um desvio de ângulo de **59$^o$,324**.

**Tabela IX – Coeficientes das retas ajustadas aos pontos para
serem corrigidos do efeito do azimute de observação solar.**

| Coeficiente | Valores a leste | Valores a oeste | Todos os valores |
|---|---|---|---|
| Linear (") | 959,130±0,053 | 958,968±0,056 | 959,065±0,038 |
| Angular ("/ °) | $0,9565.10^{-3}±8,295.10^{-4}$ | $3,830.10^{-3}±8,922.10^{-4}$ | $2,130.10^{-3}±6,066.10^{-4}$ |

Quando ajustados a uma parábola os pontos observados em função do valor absoluto do desvio do azimute de observação do Sol para a direção norte seguem uma expressão da forma: **Y = A + B1.X + B2.X$^2$**, onde **Y** é o semidiâmetro do Sol e **X** o desvio do azimute. **A** é o coeficiente independente, **B1** é o coeficiente de primeiro grau e **B2** o coeficiente de segundo grau. A Tabela X mostra os coeficientes das parábolas ajustadas aos pontos a leste, aos pontos a oeste e a todos os pontos tomados globalmente.

**Tabela X – Coeficientes das parábolas ajustadas aos pontos para
serem corrigidos do efeito do azimute de observação solar.**

| Coeficientes | Valores a leste | Valores a oeste | Todos os valores |
|---|---|---|---|
| Independente (") | 957,853±0,164 | 959,132±0,179 | 958,412±0,120 |
| Primeiro grau ("/o) | $4,838.10^{-2}±5,840.10^{-3}$ | $-0,2000.10^{-2}±6,130.10^{-3}$ | $2,587.10^{-2}±4,190.10^{-3}$ |
| Segundo grau ("/o$^2$) | $-3,746.10^{-4}±4,569.10^{-5}$ | $4,625.10^{-5}±4,811.10^{-5}$ | $-1,879.10^{-4}±3,282.10^{-5}$ |

Na Figura XXII pode-se ver os pontos observados a leste e os seu ajuste a uma reta e a uma parábola e na Figura XXIII os pontos observados a oeste e o seu ajuste a uma reta e a uma parábola. Na Figura XXIV pode-se ver os todos os pontos observados e o seu ajuste a uma reta e para uma parábola.

Comparando-se os dois lados vemos que num ajuste linear, ambos são crescentes em função do desvio azimutal, embora o desvio dos pontos a oeste seja mais forte. No ajuste de segundo grau temos diferenças maiores, enquanto a leste a concavidade da parábola é para baixo e bem pronunciada, a oeste se dá o contrário, isto é, a concavidade é para cima e bem menos pronunciada. A reta ajustada aos pontos tomados globalmente, naturalmente, tem valores crescentes, a uma taxa intermediaria entre as taxas de crescimento das curvas a leste e a oeste. A parábola tem a concavidade para baixo, uma vez que a parábola ajustada a leste é mais pronunciada que a parábola ajustada a oeste.

**Figura XXII - Os valores corrigidos de semidiâmetro solar, observados a leste, em função dos desvios do azimute do Sol para a direção Norte. Os pontos, e seus ajustes a uma reta e a uma parábola.**

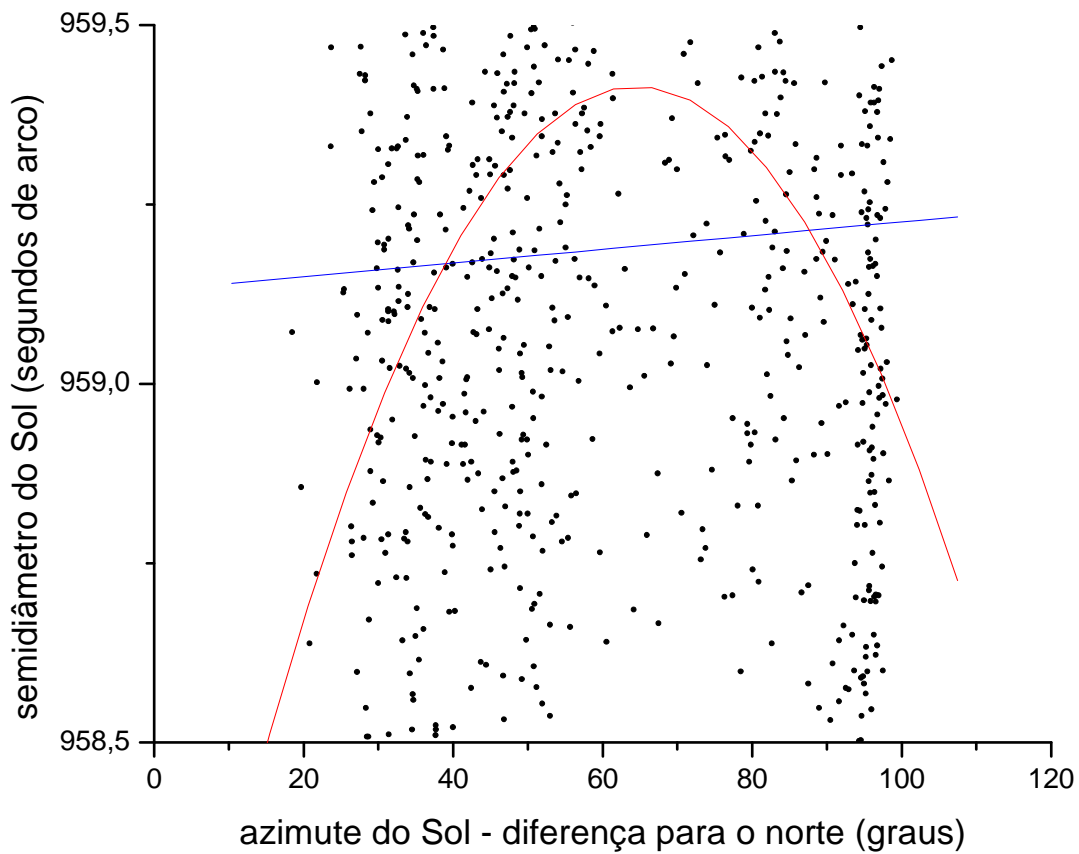

O ajuste a retas crescentes no mesmo sentido revela uma tendência natural dos pontos observados, já que o aumento de ângulos em relação à direção norte está associado à evolução temporal, do inverno para o verão, enquanto que o ajuste a parábolas com concavidades opostas revela alguma influência dos azimutes de observação solar, isto é, de algum desnivelamento do instrumento, nos valores medidos. Assim considerando, decidimos corrigir os pontos observados retirando-lhes a tendência a se ajustarem a cada uma das parábolas acima descritas, sem retirar, por outro lado, sua tendência às retas.

**Figura XXIII - Os valores corrigidos de semidiâmetro solar, observados a oeste, em função dos desvios do azimute do Sol para a direção Norte. Os pontos, e seus ajustes a uma reta e a uma parábola.**

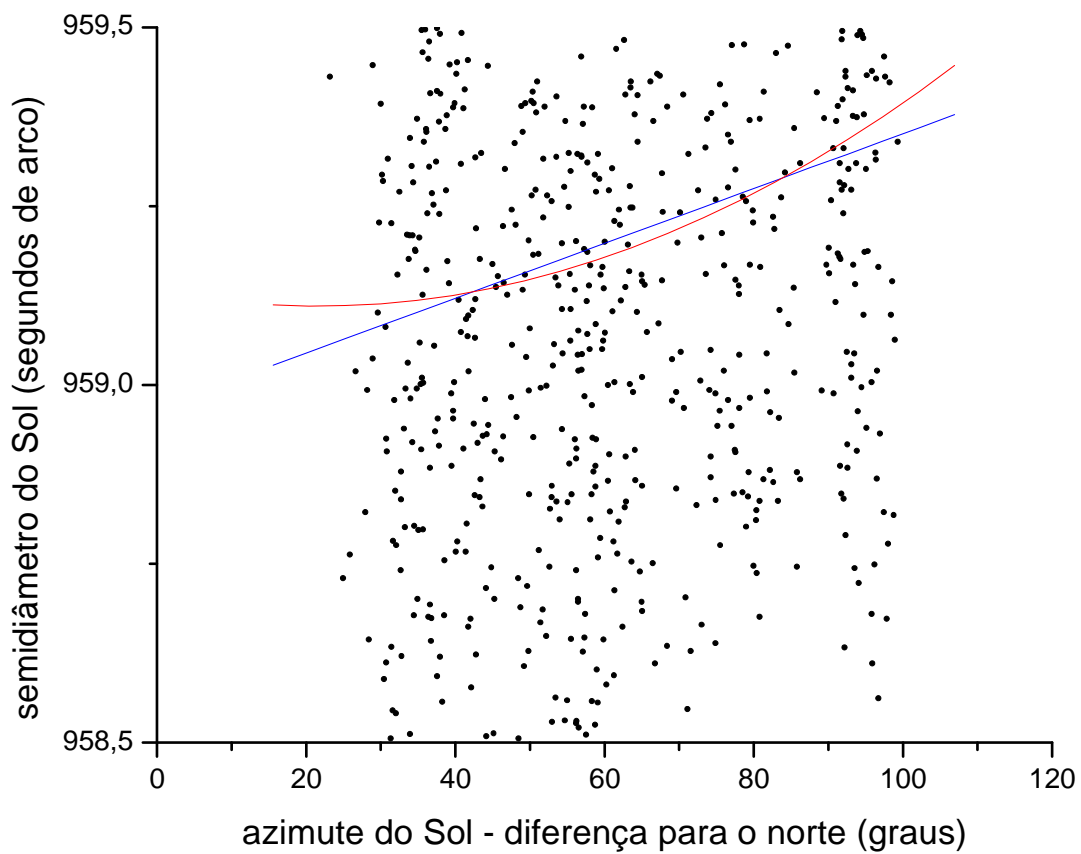

Para corrigir a tendência dos pontos a se ajustarem a uma parábola sem contudo, retirar a tendência de se ajustarem a uma reta procedemos da seguinte forma: Criamos uma nova série de pontos corrigidos da tendência linear. A esta nova série ajustamos uma parábola. Os pontos da série anterior devem ser corrigidos de acordo com os parâmetros desta nova parábola para manterem sua tendência linear. Este procedimento é feito para cada um dos lados, leste e oeste.

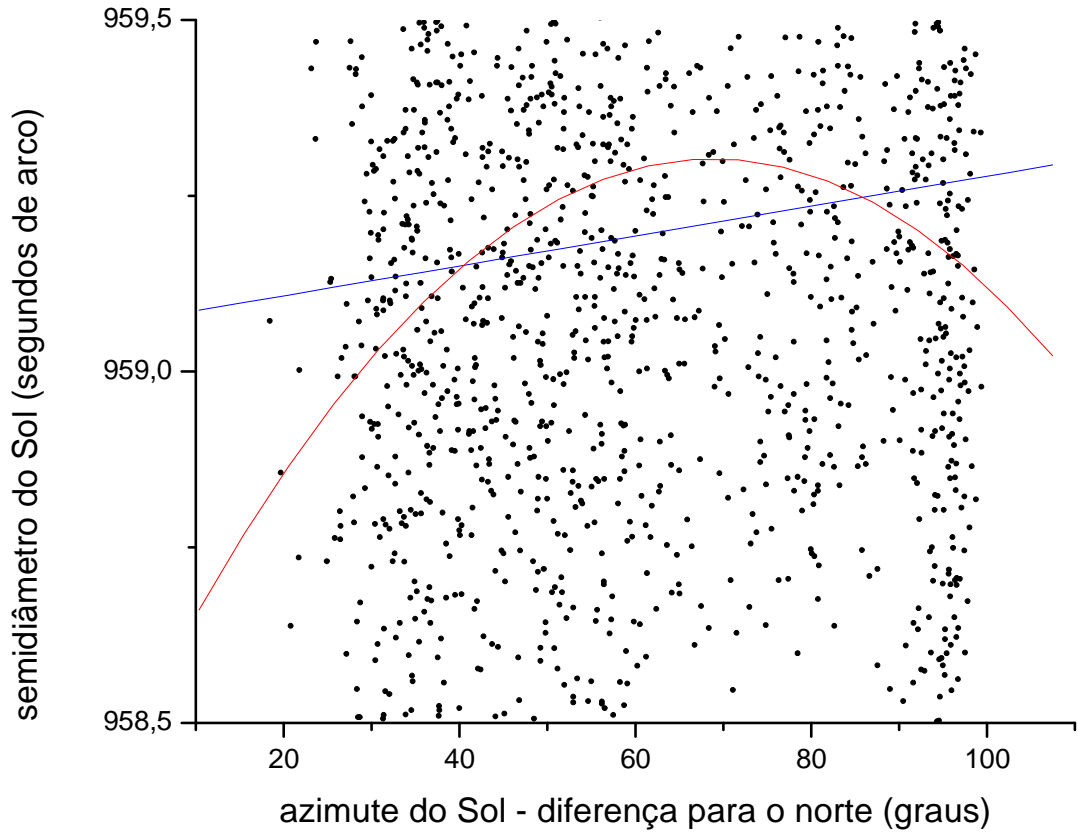

**Figura XXIV - Os valores corrigidos de semidiâmetro solar em função dos desvios do azimute do Sol para a direção Norte. Todos os pontos e seus ajustes a uma reta e a uma parábola.**

Para produzir a série sem tendência linear os valores a leste foram somados a **9,56485.10$^{-4}$.(58,589523 – X)**, e os valores a oeste foram somados a **0,00383.(59,32376 – X)**, onde **X** é o módulo da diferença do azimute de observação do Sol para a direção Norte. O número que é subtraído de **X** é o valor de **X** onde a reta que se ajusta aos pontos corta o valor médio. O fator multiplicativo é o parâmetro angular desta reta.

As parábolas ajustadas a estas séries, sem tendência linear, tem a forma **Y = A+B1.X+B2.X$^2$**, onde **Y** é o semidiâmetro solar e **X** é o módulo da diferença do azimute de observação do Sol para a direção Norte. **A** é o coeficiente independente, **B1** é o coeficiente de primeiro grau e **B2** o coeficiente de segundo grau. A Tabela XI mostra os coeficientes para as duas parábolas, aquela ajustada aos pontos a leste e a outra, ajustada aos pontos a oeste.

**Tabela XI – Coeficiente das parábolas ajustadas à série de pontos sem tendência linear para serem corrigidos do efeito do azimute de observação solar.**

| Coeficientes | Valores a leste | Valores a oeste |
|---|---|---|
| Independente (") | 957,909±0,164 | 959,359±0,179 |
| Primeiro grau ("/o) | $4,742.10^{-2} \pm 5,840.10^{-3}$ | $-0,5830.10^{-2} \pm 6,130.10^{-3}$ |
| Segundo grau ("/o$^2$) | $3,746.10^{-4} \pm 4,569.10^{-5}$ | $0,4625.10^{-4} \pm 4,811.10^{-5}$ |

A correção sofrida pelos pontos observados a cada lado foi no sentido de se subtrair destes o valor $\Delta Y = C + B1.X + B2.X^2$ onde $\Delta Y$ é a correção em segundos de arco a ser subtraída do semidiâmetro solar e $X$ é o módulo do desvio do azimute de observação solar para a direção Norte. Os dois fatores que multiplicam $X^2$ e $X$, são os coeficientes das parábolas ajustadas às séries sem tendência linear, e o termo independente $C$ é a diferença entre o termo independente destas parábolas $A$ e a média dos pontos observados a cada lado. Para o lado leste $C = -1,2775$ e para o lado oeste $C = 0,16387$ segundos de arco.

As correções assim aplicadas foram de um mínimo de **–0,072** a um máximo de **0,020** segundos de arco sendo a média igual a **–0,004** segundos de arco e o desvio padrão igual a **0,019** segundos de arco. A Figura XXV mostra um histograma da distribuição das correções. A distribuição das correções tem um valor sempre crescente do mínimo em direção ao máximo, isto é, em torno de uma correção de –0,07 segundos de arco há um número pequeno de correções feitas, mas em torno de 0,02 segundos de arco há um grande número de correções.

Com estas correções a média dos pontos a leste permaneceu igual a **959,186** segundos de arco e o desvio padrão caiu de **0,621** para **0,601** segundos de arco. A média dos pontos a oeste permaneceu **959,195** segundos de arco e o desvio padrão caiu ligeiramente de **0,558** para **0,557** segundos de arco. Tomados os pontos, globalmente, sua média passou de **959,191** para **959,190** segundos de arco e o desvio padrão caiu de **0,591** para **0,580** segundos de arco. Como era de se esperar, ao se retirar a influência externa aos valores medidos, o desvio padrão caiu, melhorando a qualidade do conjunto.

**Figura XXV - Histograma das correções aplicadas aos valores observados de semi-diâmetro do Sol para corrigir erros introduzidos pelo azimute de observação do Sol.**

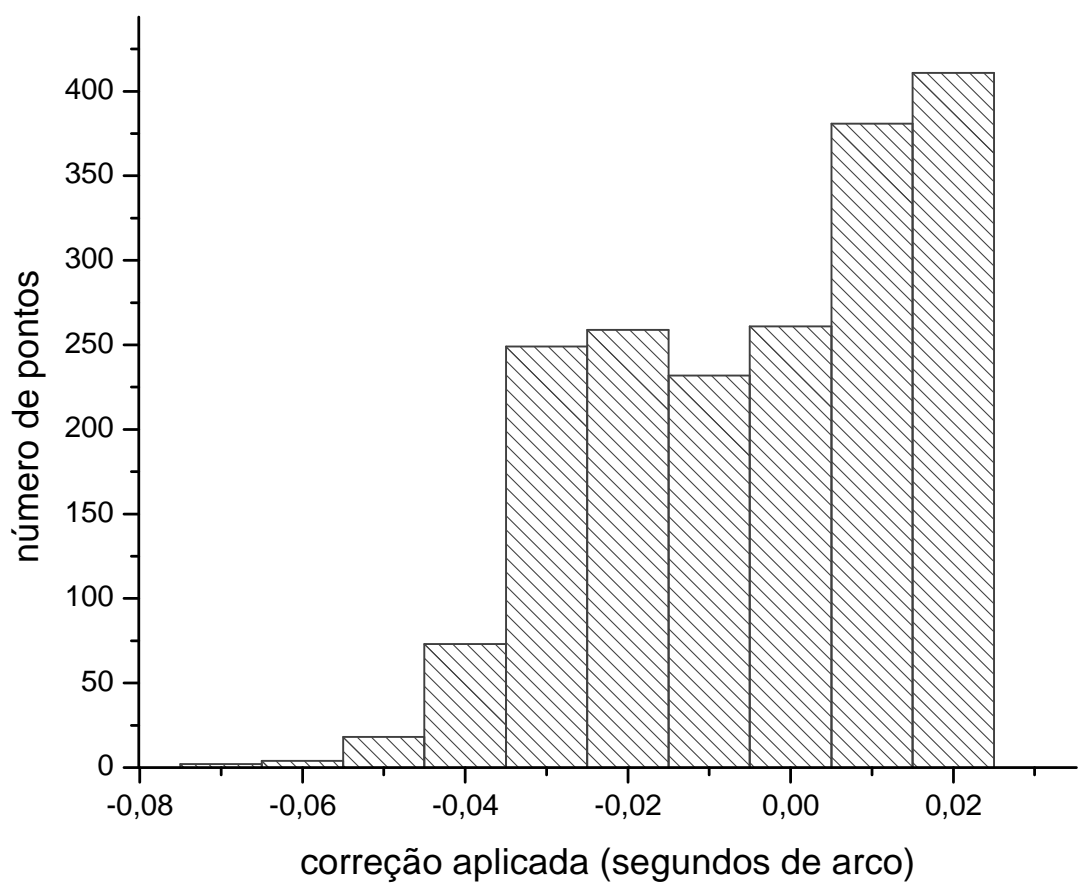

**A SÉRIE FINAL.**

Com estas, esgotamos o elenco de correções que devíamos efetuar na série original de dados. Temos agora, uma nova série de dados observados e corrigidos que é apresentada na Figura XXVI. Nela aparecem os valores corrigidos de semidiâmetro solar em uma série temporal em função da data Juliana modificada. Para poder se ver a evolução média dos pontos, foi traçada uma média móvel de 100 pontos para a série de pontos observados a leste e para a série de pontos observados a oeste.

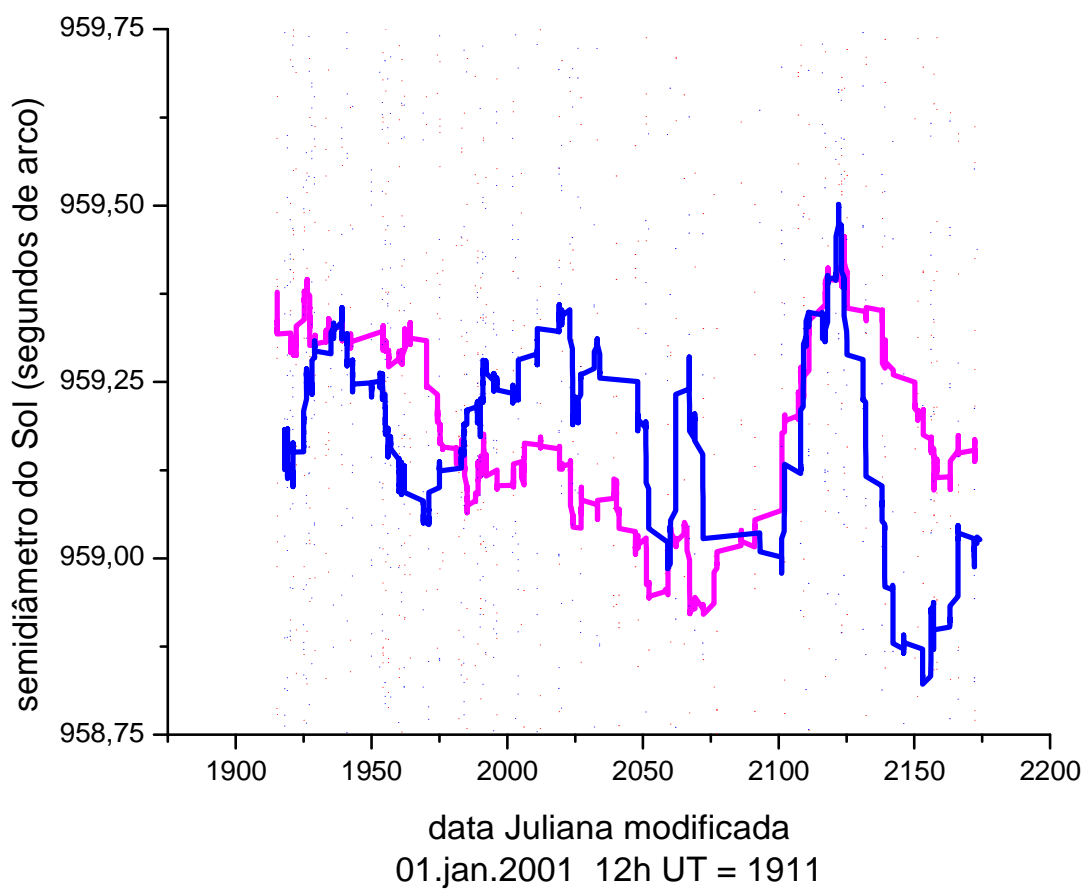

**Figura XXVI - Valores corrigidos das observações do semidiâmetro do Sol em duas séries: observados a leste em azul e a oeste em magenta. As curvas representam as médias móveis a cada cem pontos.**

A nossa série de pontos têm média de **959,190** segundos de arco e desvio padrão de **0,580** segundos de arco. Sendo a média dos pontos a leste têm média de **959,186** segundos de arco e desvio padrão de **0,601** segundos de arco. A média dos pontos a oeste é de **959,195** segundos de arco e desvio padrão de **0,557** segundos de arco. O histograma da distribuição dos pontos pode ser visto na Figura XXVII. A esta distribuição está ajustada uma Gaussiana e o teste do Chi-quadrado forneceu um valor de 50,9213, sendo que para 21 graus de liberdade, como é o caso, um valor superior a 41,401 indica que os pontos seguem uma distribuição normal com 99,5% de certeza. Esta normalidade aparece por conta dos intervalos escolhidos, já que na realidade não há um valor médio, mas uma tendência em torno da média.

**Figura XXVII - Histograma das observações do semidiâmetro solar depois de todas as correções. Distribuídos em 21 intervalos de 0,2 segundos de arco, desde 957,0 até 961,2 segundos de arco**

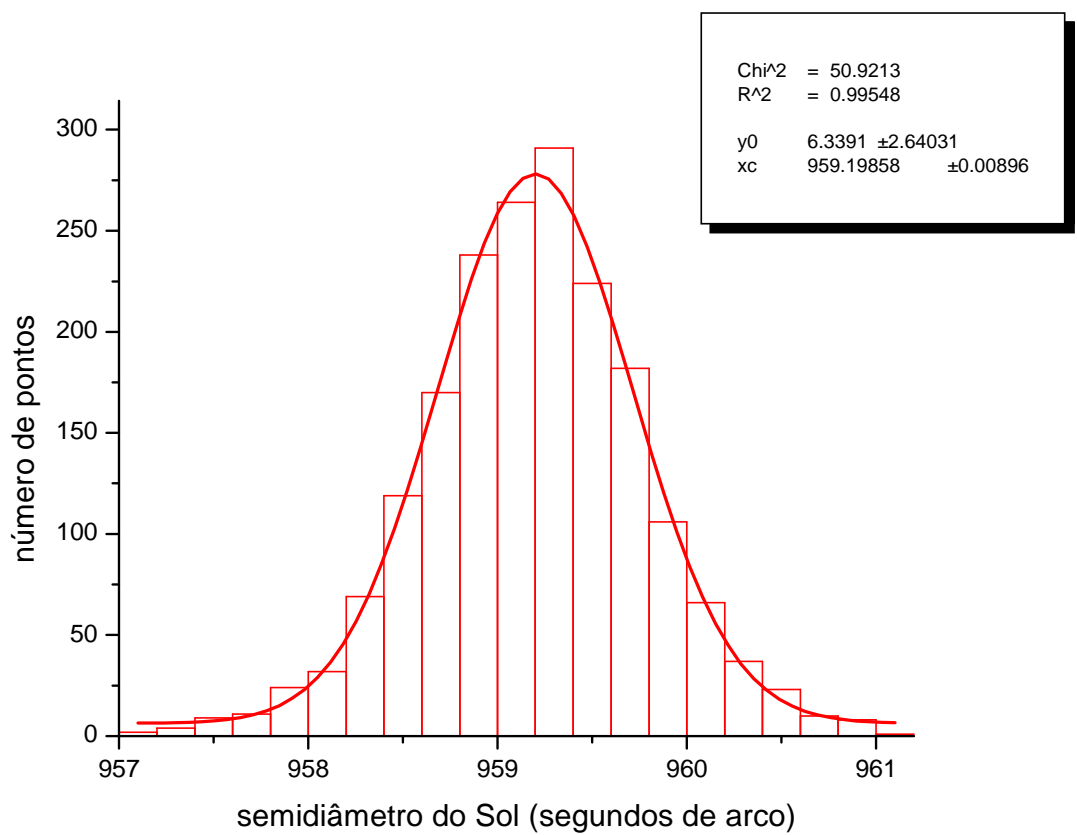

A Tabela XII, que segue mostra a evolução da média e dos desvios-padrão, para dados observados a leste, a oeste e para o seu total, em quatro etapas do nosso trabalho. Inicialmente (Inic.), após a correção causada pela instabilidade de posicionamento do prisma (Pris.), após a correção da influência de parâmetros de observação (Para.), e após a correção pelo azimute de observação solar (Azim.).

**Tabela XII – Evolução da média e do desvio padrão ao longo do processo de correção dos valores observados de diâmetro do Sol.**
**(valores em segundos de arco)**

| Etapas | LESTE | | OESTE | | TOTAL | |
|---|---|---|---|---|---|---|
| | Média | Desvio | Média | Desvio | Média | Desvio |
| (Inic.) | 959,259 | 0,643 | 959,115 | 0,564 | 959,189 | 0,610 |
| (Pris.) | 959,191 | 0,628 | 959,190 | 0,556 | 959,191 | 0,594 |
| (Para.) | 959,186 | 0,621 | 959,195 | 0,558 | 959,191 | 0,591 |
| (Azim.) | 959,186 | 0,601 | 959,195 | 0,557 | 959,190 | 0,580 |

Esta tabela é retratada nas Figuras XXVIII, XXIX e XXX, que mostram a evolução das médias e dos desvios-padrão ao longo do processo de correção dos valores observados. Fica evidente que as observações a leste são as que sofrem as maiores correções e têm o seu desvio-padrão bastante diminuído. Os desvios-padrão das observações a oeste são a todo tempo menores que os desvios-padrão das observações a leste, mesmo o menor valor, após a última correção. As correções aplicadas aos valores observados podem ser vistas na Tabela XIII que se segue.

**Tabela XIII – Mudanças na média e no desvio padrão em função das correções aplicadas.**
**(valores em segundos de arco)**

| correções | LESTE | | OESTE | | TOTAL | |
|---|---|---|---|---|---|---|
| | Média | Desvio | Média | Desvio | Média | Desvio |
| (1ª corr.) | -0,0679 | -0,0156 | 0,0755 | -0,0082 | 0,0015 | -0,0164 |
| (2ª corr.) | -0,0045 | -0,0069 | 0,0048 | 0,0018 | 0,0000 | -0,0029 |
| (3ª corr.) | 0,0000 | -0,0203 | -0,0002 | -0,0003 | -0,0001 | -0,0111 |

Esta tabela mostra que a correção devida à instabilidade do prisma objetivo destaca-se como a correção de maior peso. Nela, os valores a leste tiveram a média reduzida e os valores a oeste tiveram a média aumentada, ambos, de valores pouco maiores que um décimo do desvio padrão. O desvio padrão foi reduzido em cerca de 2,5% de seu valor a leste em pouco mais de 1% a oeste, resultando numa diminuição de pouco mais de 2,5% para o total dos pontos

**Figura XXVIII - Evolução das médias das observações
ao longo dos processos de correção.**

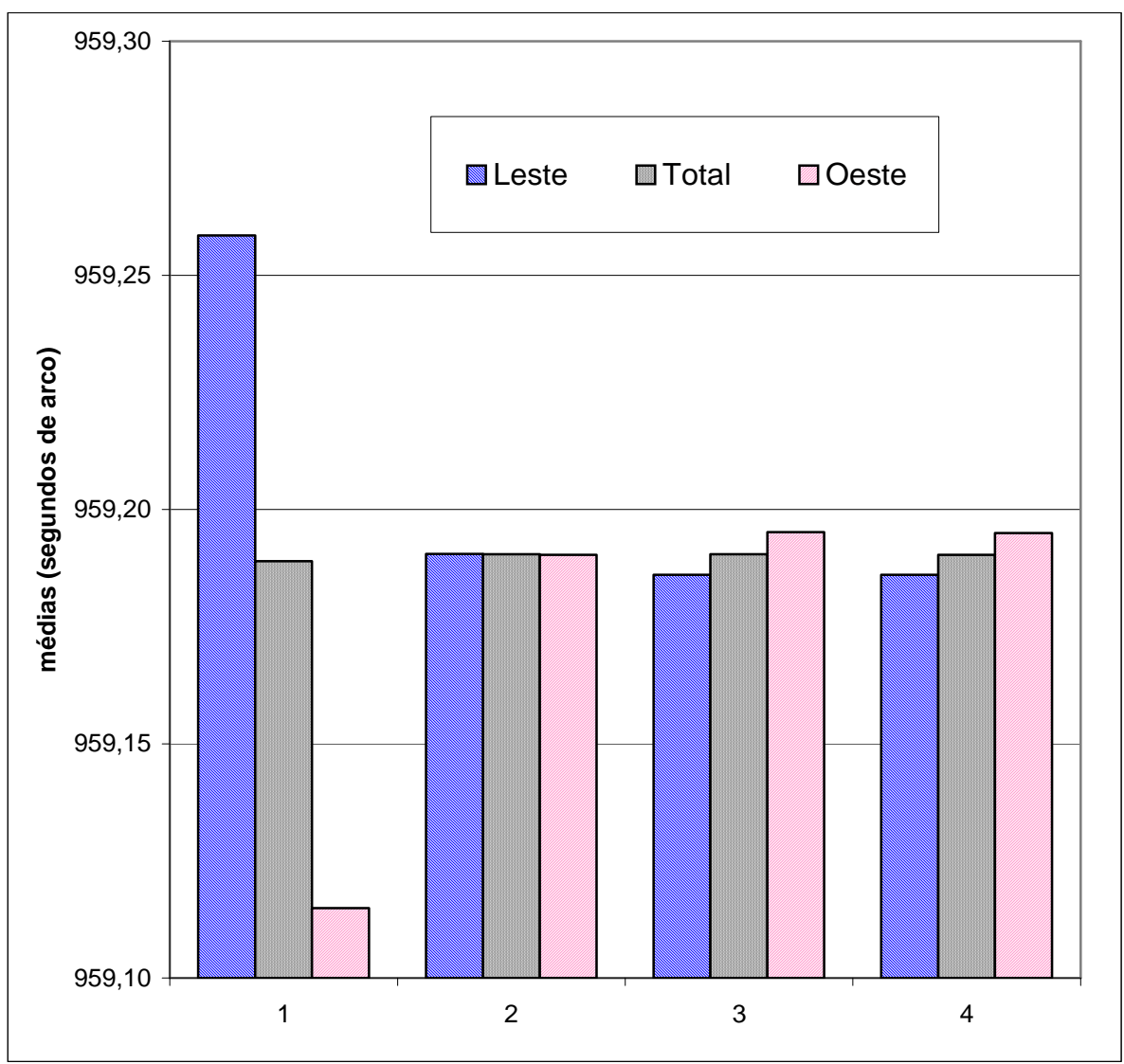

A correção devida aos parâmetros das observações foi bem menor. A média dos valores a leste foi diminuída e a dos valores a oeste aumentada, ambas em cerca de 0,75% do valor do desvio padrão. O desvio padrão a leste foi diminuído em mais de 1% de seu valor, enquanto que a oeste ele teve um pequeno aumento de cerca de 0,3%, resultando em uma diminuição de 0,5% de seu valor para o total dos pontos.

**Figura XXIX - Evolução das médias das observações ao longo dos processos de correção – apenas as três etapas finais.**

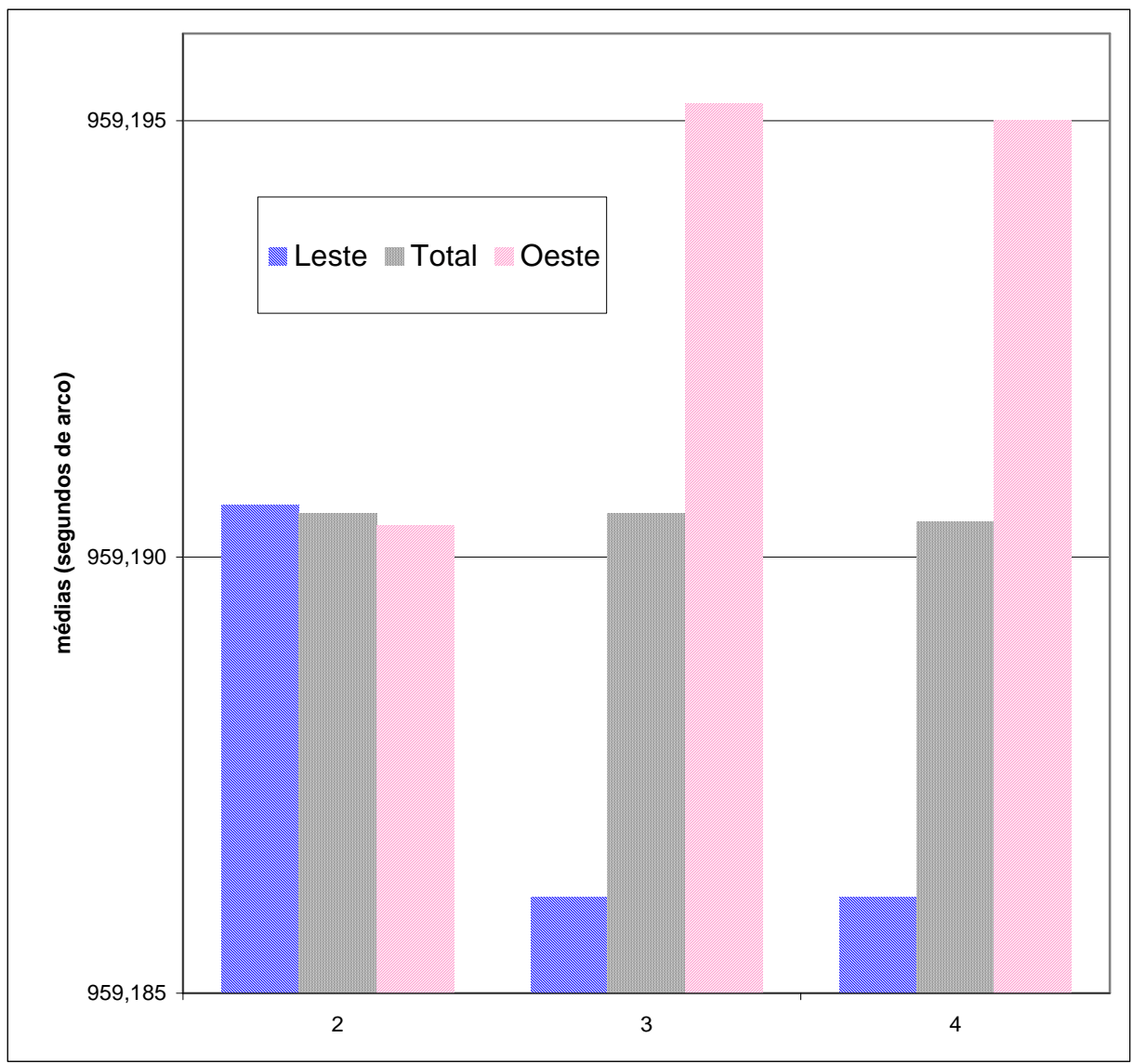

A correção devida ao azimute de observação do Sol praticamente não afetou as médias dos valores, nem a leste, nem a oeste, nem tampouco afetou o desvio padrão dos valores a oeste, mas foi a correção que mais reduziu o desvio padrão dos valores observados a leste, ele foi reduzido em quase 3,5% de seu valor anterior, resultando em uma redução de quase 2% para o total dos valores.

**Figura XXX - Evolução dos desvios-padrão das observações ao longo dos processos de correção.**

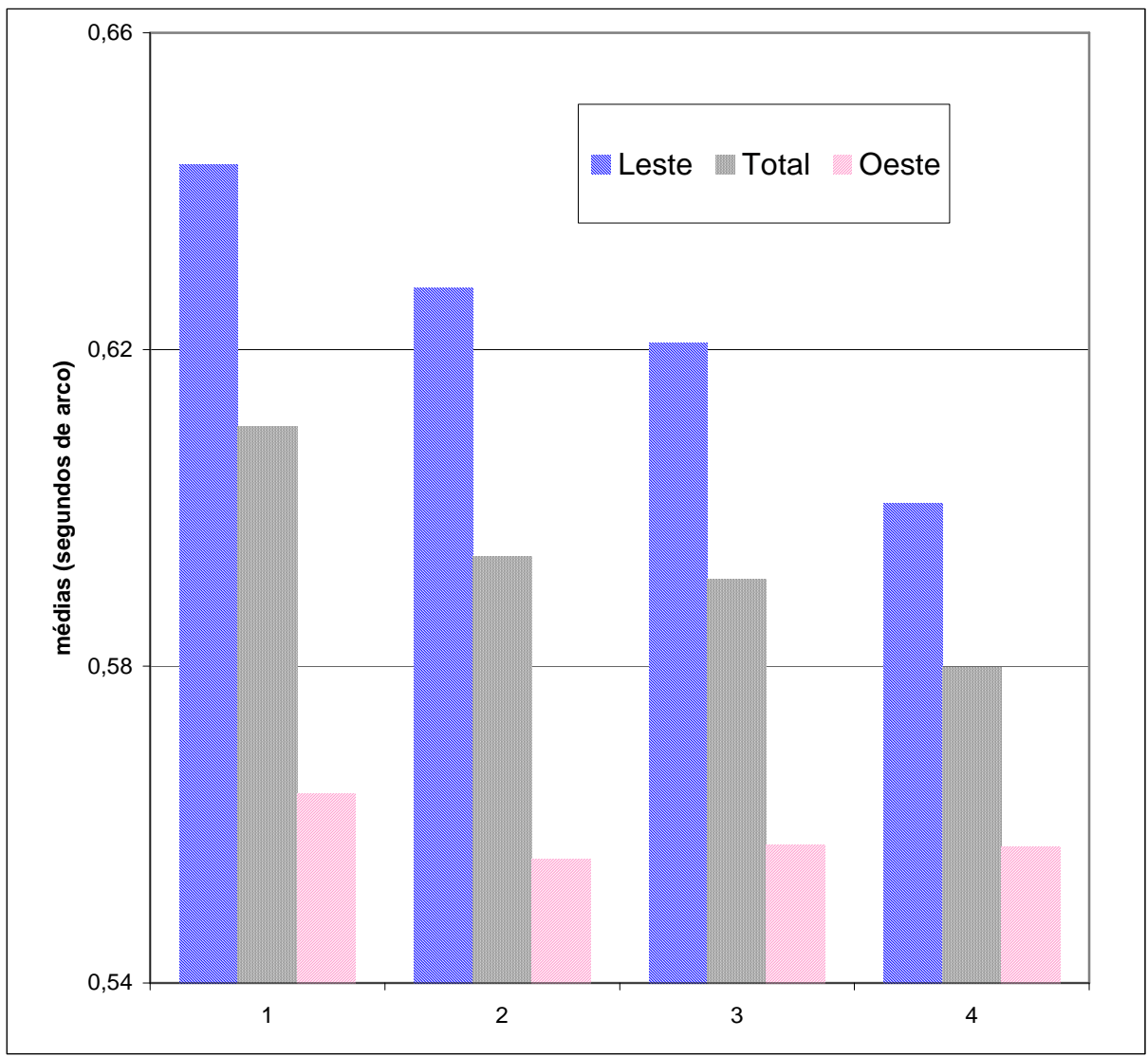

# COMPARAÇÃO COM OUTRAS CURVAS.

## Com Outros Observatórios.

Uma vez disponível a série de pontos observados em 2001 e corrigidos dos erros introduzidos por efeitos de instrumento e de parâmetros de observação podemos compará-los com outras séries produzidas por outras equipes que também observam o semidiâmetro solar.

Este é o caso da série produzidas pelo CERGA – Centre d'études et de Recherches em Géodynamique et Astrométrie, [13] que é um departamento do Observatório da Cote d'Azur na França. Temos também a série obtida pelo Observatório de Tubitak em Antalya na Turquia [14].

A série francesa têm pouco mais de 1300 observações. Há problemas de observação, principalmente no inverno, de modo que a série inicia apenas em 14 de março e termina em 18 de outubro e tem um intervalo entre 3 de abril e 25 de maio. A série da Turquia é menor, tem cerca de 550 observações, isto porque seu astrolábio tem três prismas fixos o que só permite a observação a três alturas zenitais. A série de Antalya se estende de 23 de fevereiro a 7 de novembro.

Não está no escopo deste trabalho fazer uma análise quantitativa da semelhança das três séries. Faremos apenas uma discussão qualitativa breve, apontando algumas semelhanças entre as três curvas em uma inspeção visual. Para tal montamos a Figura XXXI onde colocamos as três séries alisadas por um método que utiliza séries de Fourier, cortando as altas freqüências. Nela, as três séries foram calculadas utilizando-se o mesmo critério de corte. Nesta figura, inicialmente nos chama a atenção o fato da série do CERGA estar deslocada para cima. Isto não se configura como uma diferença, pois, além do semidiâmetro do Sol ser variável com o comprimento de onda observado, depende também de como se ajusta o bordo solar observado pelo astrolábio. O próprio ON, como já dissemos, determinou três curvas diferentes de semidiâmetro solar em função do desvio padrão adotado para os pontos que ajustados configuravam o bordo solar. As três curvas, bastante semelhantes, diferem apenas por uma diferença constante de semidiâmetro. Assim, podemos comparar as séries admitindo sempre um deslocamento vertical constante no gráfico para toda uma série.

**Figura XXXI - Semidiâmetros do Sol observados durante o ano de 2001, no ON em verde, no CERGA em azul e em Antalya em vermelho.**

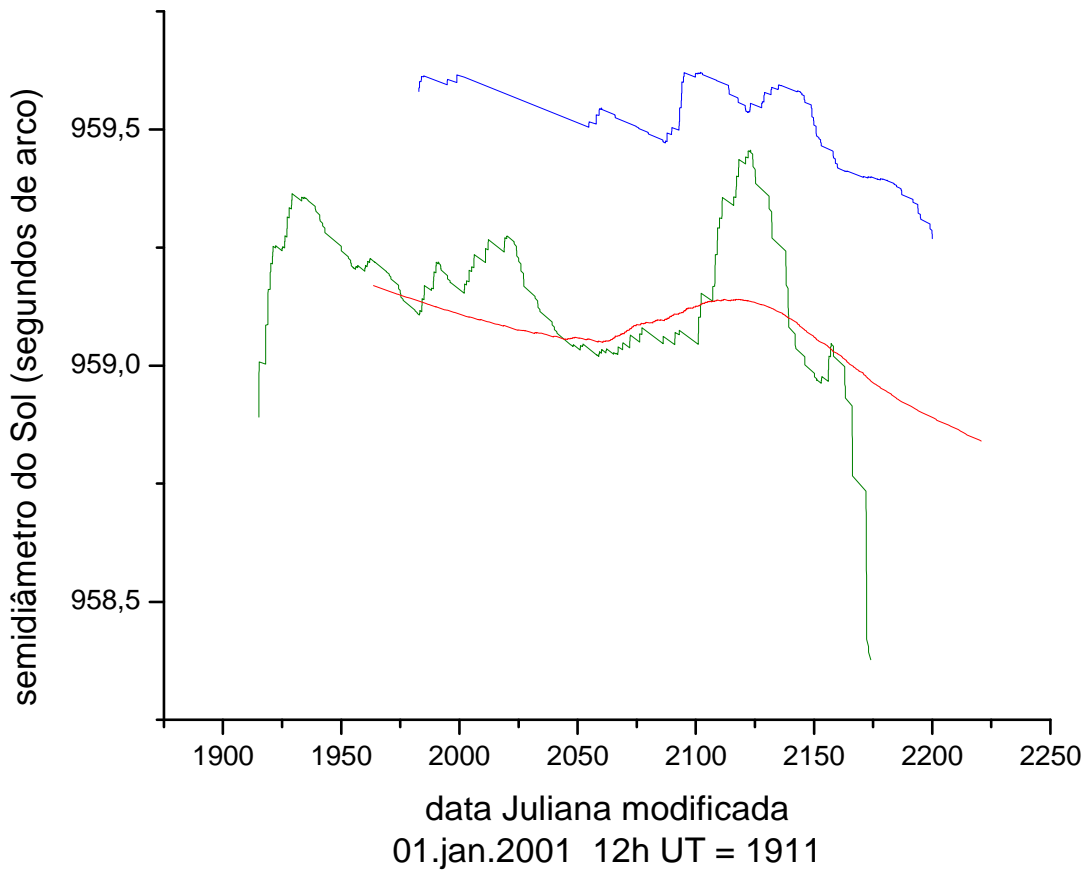

Devemos não considerar nesta figura os extremos das curvas traçadas que são distorcidos justamente por não contarem com mais pontos para além ou aquém dos extremos. Isto ocorre principalmente com a curva do ON e em menor grau com a curva do CERGA. Dito isto podemos, de uma maneira qualitativa, observar, inicialmente, uma mesma tendência temporal nas três curvas. Todas têm uma tendência decrescente cujo ângulo de inclinação é semelhante.

Se desconsiderarmos na curva francesa o período entre 3 de abril e 25 de maio, que têm data Juliana modificada, aproximadamente, entre 2000 e 2050, no qual não houve observações, então, esta curva e a curva do ON guardam muita semelhança. Há, entretanto, naquela curva dois picos de máximos separados por um vale, enquanto que na outra há apenas um máximo.

Este vale está relacionado ao desequilíbrio de observações a leste e a oeste, justamente neste intervalo. Ao considerarmos este fato as duas curvas, qualitativamente, se aproximam ainda mais.

O vale no meio do máximo da curva do CERGA é explicado por uma distribuição local não usual de observações a leste e a oeste. Para toda série a proporção é 57,5% observações a leste, para o vale a proporção é 47,5% observações a leste. À esquerda do vale a proporção é 62,0% observações a leste, à direita do vale a proporção é 62,5% observações a leste. Tomando separadamente (leste ou oeste) o vale, na verdade, é um máximo. Ele é 0",03 maior que a média dos valores laterais, tanto a leste como a oeste. O vale aparece porque nele a proporção de observações a oeste é bem maior (52,5%, contra 40,0% na média).

A curva de Antalya parece mais alisada que as outras, isto é uma decorrência de um número mais reduzido de observações, mas ainda assim, mostra também o mesmo máximo que as duas outras curvas no início de agosto, ou seja, para a data Juliana modificada em torno de 2125.

**Com a Atividade Solar.**

Podemos comparar nossa série com a atividade solar. Aqui, também, não é escopo de nosso trabalho determinar estatisticamente a semelhança das duas séries, mas, fazer uma breve avaliação visual dos gráficos e apontar possíveis semelhanças. Para tal produzimos a Figura XXXII, onde temos a nossa série atrasada de 50 dias e o número médio diário de manchas solares em uma escala arbitrária. Esta escala foi assim concebida: tomamos os valores, retiramos deles o valor médio e dividimos o resultado por 180. Este valor é um número que torna o desvio padrão da série de manchas solares comparável com o desvio padrão da nossa série de observações do semidiâmetro do Sol. Ao valor então obtido somamos 958,7, de modo que as duas séries ocupassem a mesma janela vertical de espaço no gráfico, sem se justaporem demais para que pudéssemos compará-las visualmente.

Embora a tendência temporal das duas séries seja oposta, ou seja, a série de semidiâmetro do Sol é decrescente, enquanto a série de número de manchas é crescente, a inspeção visual das duas curvas sugere que a série de semidiâmetros do Sol, atrasada de 50 dias, tem seus máximos bastante coincidentes com os máximos de números de manchas solares. A diferença de 50 dias entre os máximos das duas curvas poderia ser justificada por algum processo com retardo no tempo.

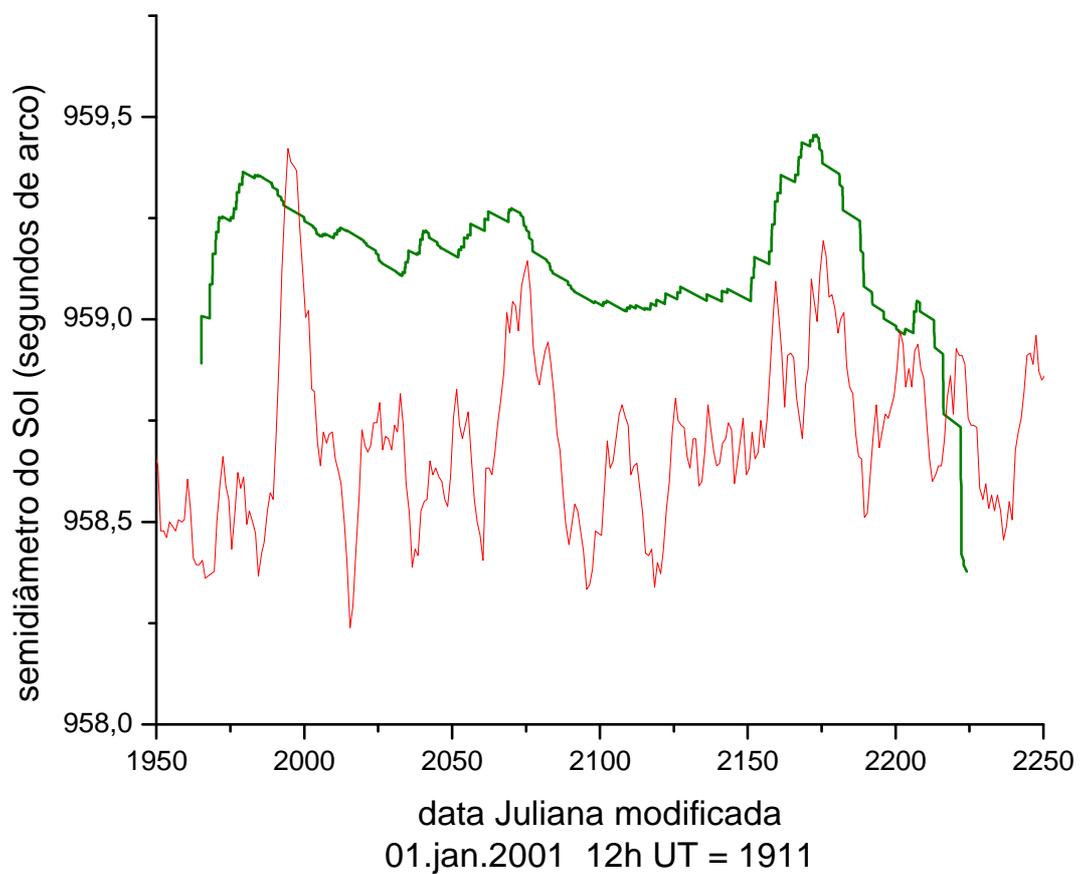

**Figura XXXII - Semidiâmetros do Sol observados durante o ano de 2001 no ON em verde, comparados com o número médio diário de manchas solares no mesmo período em uma escala arbitrária, em vermelho.**

# CONCLUSÕES.

Os dados de observações do semidiâmetro do Sol feitas no ON durante o ano de 2001, foram tratados de modo a se retirar deles influências externas introduzidas por alguns fatores ligados ao instrumento de observação ou ao modo como dele se utiliza. Ao se retirar estas influências, obtivemos uma série de valores mais adequados para a análise, bem como, para serem posteriormente utilizados em outras investigações.

Os erros instrumentais são muito maiores que os erros observacionais o que concorre para se admitir que a qualidade observacional é bastante boa. Assim podemos afirmar que os efeitos da atmosfera não chegam a ser relevantes na qualidade das medidas.

As correções encontradas foram sempre bem inferiores ao desvio padrão dos valores observados, a correção é um pouco superior a um sétimo do desvio padrão a oeste e um pouco inferior a um oitavo a leste. Isto garante a integridade dos dados observados.

A série final corrigida aponta claramente para uma variação linear do semidiâmetro solar durante o ano de 2001. Levando a uma diminuição da ordem de décimos de segundo de arco no ano. Além desta, há claramente variações locais como um pico de máximo no início de agosto, em torno da data Juliana modificada igual a 2125. O pico em seu ponto máximo se destaca da variação linear por quase meio segundo de arco.

A variação mais constante apontada é da ordem de 0,01% e as variações locais são da ordem de 0,05%. Isto as torna compatíveis com as variações de fluxo solar que, neste mesmo período, são da mesma ordem de grandeza.

Os dados corrigidos são comparáveis com as séries observadas em outras latitudes. Uma inspeção visual das séries sugere uma boa semelhança entre elas. Embora não seja conclusiva, a análise mostra não haverem discrepâncias entre as três séries comparadas. Isto indica, mais uma vez, que os efeitos atmosféricos não são relevantes na medida.

Observar o Sol de diferentes latitudes significa observar o seu semidiâmetro em diferentes latitudes solares. Não procede a questão de que a observação de diferentes diâmetros solares pode levar a resultados incompatíveis pela presença ou ausência de manchas no limbo solar.

Merece investigação maior a comparação da série com a atividade solar, pois um exame visual das duas curvas mostra semelhanças desde que se atrase a curva de semidiâmetro do Sol de um valor adequado. Aqui, entretanto, há a necessidade de um período maior de dados para uma comparação mais efetiva, bem como o uso de outros parâmetros da atividade solar.